\documentclass[11pt]{article}  \usepackage{geometry}

\usepackage{natbib}
\def\citeN{\citet}

\usepackage{xspace} 
\def\url{url~}

\usepackage{amsmath,amssymb,listings}

\usepackage{caption,float}

\usepackage{epsfig,graphicx,epstopdf,eurosym} 
\usepackage{tikz}
\usetikzlibrary{calc,positioning, shapes, fit,arrows}
\usepackage{multirow,cases}
\usepackage{eurosym}
\usepackage[labelfont=bf]{caption}
\usepackage{hyperref}

\usepackage{pgf} 
\usepackage{array,color,colortbl}
\usepackage{mathrsfs}
\usepackage{adjustbox}
\definecolor{midnightblue}{rgb}{0.10,0.10,0.44}
\definecolor{rltgreen}{rgb}{0,0.4,0}
\definecolor{rltred}{rgb}{0.75,0,0}
\definecolor{lightgreen}{rgb}{0,0.6,0}
\definecolor{LemonChiffon}{rgb}{1.,0.98,0.8}
\colorlet{mauve}{blue!70!red}

\usepackage{dsfont}
\definecolor{WildStrawberry}{rgb}{1.00,0.00,0.50}
\definecolor{ForestGreen}{rgb}{0.00,1.00,0.00}
\definecolor{Blue}{rgb}{0.00,0.00,1.00}
\definecolor{WildStrawberry}{rgb}{0.5,0.00,0.25}
\definecolor{ForestGreen}{rgb}{0.00,50,0.50}
\definecolor{Blue}{rgb}{0.00,0.00,1.00}

\def\tilde{\widetilde}

\def\ben{\begin{enumerate}}
\def\een{\end{enumerate}}
\def\bit{\begin{itemize}}
\def\eit{\end{itemize}}
\def\bqo{\begin{quotation}}
\def\eqo{\end{quotation}}
\def\bdi{\begin{description}}
\def\edi{\end{description}} 
\def\bb{\begin{block}}
\def\eb{\end{block}} 

\def\bal{\begin{aligned}}
\def\eal{\end{aligned}}

\def\proof{\noindent {\it Proof. $\, $}}\def\proof{\noindent{\it\textbf{Proof}. $\, $}}
\def\finproof {\hfill $\Box$ \vskip 5 pt }

\def\bdm{\begin{displaymath}}
\def\edm{\end{displaymath}}
\def\be{\begin{enumerate}}
\def\ee{\end{enumerate}}
\newcommand{\beqa}{\begin{eqnarray}}
\newcommand{\eeqa}{\end{eqnarray}}
\newcommand{\beqe}{\beqa\begin{aligned}}
\newcommand{\eeqe}{\end{aligned}\eeqa}

\def\bea{\begin{eqnarray*}}
\def\eea{\end{eqnarray*}}
\def\bi{\vskip -7 pt\begin{itemize}\vskip -7 pt}
\def\ei{\vskip -7 pt\end{itemize}\vskip -7 pt}

\newcommand{\bde}{\begin{displaymath}}
\newcommand{\ede}{\end{displaymath}}
\newcommand{\bel }{\left\{\begin{array}{ll}}
\newcommand{\eel}{\cr \end{array} \right.}

\def\t{\tilde}

\def\cL{\mathcal{L}}

\def\cJ{{\cal J}}

\def\cF{{\cal F}}

\def\cP{{\cal P}}

\def\cC{{\cal C}}

\def\cL{{\cal L}}

\def\b{\beta}

\def\ca{{\check{a}}}

\def\ff{{\tk\in{\cal F}}}

\def\cdots{\ldots}

\def\eee{\end{document}}
\def\eef{\end{frame}\end{document}}

\renewcommand\be{\begin{equation}}
\renewcommand{\ee}{\end{equation}} 

\def\Xi{Y}

\def\cL{L}

\def\cJ{{\cal J}}
\def\cP{{\cal P}}

\def\cP{{\cal P}}

\def\cY{{\cal Y}}

\def\cW{{\cal W}}

\def\tilde{\widetilde}

\def\t0{0}
\def\0{0}

\def\ff{{\mathbb F}}

\def\Nom{N}

\def\HH1{{\cal H}^{2}(\ff)}

\def\L2{ L^2(0,T)}

\def\b{\,^\beta\! }

\def\M2{M^2}

\def\cY{{\cal Y}}

\def\bal{\begin{aligned}}
\def\eal{\end{aligned}}

\def\cL{{\cal Z}}

\def\be{\begin{equation}}
\def\ee{\end{equation}}

\def\ff{{\mathbb F}}

\usepackage{float}
\usepackage{epsfig}
\usepackage{graphicx}
\usepackage{amsfonts}
\usepackage{stmaryrd}

\definecolor{GrapeFruit}{rgb}{0.92,0.33,0.08}
\definecolor{Wine}{rgb}{0.5,0,0.25}

\newcommand{\A}{A}

\def\t{{(2)}}

\def\t{{\{2\}}}

\def\xitau2{\xi_{(\tau_2)}}
\def\Ltau2{\phi(\tau_1,\tau_2)}\def\Ltau2{\tilde{\xi}(\tau_1,\tau_2)}
\def\chitau2{\chi_{(\tau_2)}}

\def\ebb{\eb\end{document}}

\def\qqq{\quad\quad\quad}

\renewcommand{\bea}{\begin{eqnarray*}}

\def\bll#1{
\beqa\label{#1}\bal
}
\def\lel{\eal\eeqa}
\newcommand{\beql}[1]{\bll{#1}}
\newcommand{\eeql}{\lel}

\newcommand{\COMM}[1]{}

\usepackage{listings}

\def\cW{\Gamma}

\def\chitau2{P_{\tau_2}}

\def\ff{{\mathbb F}}

\def\xitau2{\xi}

\def\ff{{\cal G}}

\def\CVA{\mathcal{A}}\def\CVA{\Theta}

\def\tilde{\widetilde}

\def\Rf{\mathfrak{R}}\def\Rf{\mathfrak{r}}

\def\thisR{R}\def\thisR{\rho}

\def\ff{\mathbb{F}}

\def\ff{\mathbb{F}^*}

\def\ff{{\cal F}}\def\ff{\mathbb{F}}

\def\Ep{{E_{b}}}

\def\chitau{\chi}

\renewcommand{\bel}{\bea\bal}
\renewcommand{\eel}{\end{aligned}\eea}

\def\Rf{\mathfrak{R}}\def\Rf{\mathfrak{r}}

\def\thisR{R}\def\thisR{\rho}

\def\ff{\mathbb{F}}

\def\kappa{S^{\star}}

\def\Rf{\mathfrak{R}}\def\Rf{\Rf}\def\Rf{R_{f}}\def\Rf{\bar{R}_{b}}\def\Rf{\bar{R}}

\def\thisR{R}\def\thisR{\thisR }\def\thisR{R_b}

\def\chitau{\chi}

\def\kappa{k}

\def\qr#1{\eqref{#1}}

\def\thea{a}

\def\theapim{\theapim}\def\theapim{\thea}
\def\theaim{\theaim}\def\theaim{\thea}

\def\FBSDE{\FBSDE\xspace}\def\FBSDE{BSDE\xspace}
 
\def\href#1#2{#2}

\def\L{\tilde{L}}

\def\CA {{\rm CA}\xspace}

\def\ES{R}\def\ES{\mbox{ES}}\def\ES{\ES}

\renewcommand{\finproof}{\rule{4pt}{6pt}}
\renewcommand{\bea}{\begin{eqnarray*}}
\renewcommand{\eea}{\end{eqnarray*}}

\def\KVA{\kva'}
\def\IM{{\rm IM}\xspace}\def\IM{\PIM}
\def\IMo{\PIMo}

\def\PIM{{\rm PIM}\xspace}
\def\PIMo{\overline{\rm PIM}\xspace}

\def\RIMo{\overline{\rm RIM}\xspace}
\def\M{C}\def\M{\mbox{IM}}\def\M{\PIM}

\def\CA{{\rm CA}\xspace}

\def\CVA{{\rm CVA}\xspace}

\def\cL{\mathcal{L}}

\def\UCVA{{\rm \UCVA}\xspace}\def\UCVA{{\rm CVA}\xspace}
\def\UFVA{{\rm \UFVA}\xspace}\def\UFVA{{\rm FVA}\xspace}

\def\M{\mathbb{M}}\def\M{\mathbb{P}}

\def\FTP{{\rm FTP}\xspace}

\def\cPo{\overline{\mathcal{P}}}

\def\M{\mathbb{M}}\def\M{\mathbb{P}}

\def\S{S}\def\S{J}
\def\So{\bar{J}}

\def\cL{{\cal H}}\def\cL{\mathcal{L}}\def\cL{\mathcal{S}}
\def\L2{L^2}\def\L2{\cL^2}\def\L2{\cL_2}

\def\EC{{\ES }}\def\EC{{\rm CR}}
\def\ES{R}\def\ES{\mbox{ES}}\def\ES{\ES}\def\ES{{\rm EC
}\xspace}

\def\CVA{{\rm CVA}\xspace}

\def\M{{\rm M}\xspace}

\def\UCVA{{\rm \UCVA}\xspace}\def\UCVA{{\rm CVA}\xspace}
\def\UFVA{{\rm \UFVA}\xspace}\def\UFVA{{\rm FVA}\xspace}

\def\KVA{\kva'\xspace}

\def\CVA{{\rm CVA}^\circ}

\def\UCVA{{\rm \UCVA}^\circ}\def\UCVA{{\rm CVA}^\circ}
\def\UFVA{{\rm \UFVA}^\circ}\def\UFVA{{\rm FVA}^\circ}

\def\KVA{{\rm KVA}^\circ}

\def\CA{{\rm CA}^\circ\xspace} 

\def\XVA{TRC/FVA\xspace}\def\XVA{{\rm CA}\xspace}\def\XVA{{\rm XVA}\xspace}

\def\Theta{{\rm CA}}

\def\mtm{{\rm MtM} \xspace}
\def\kva{{\rm KVA} \xspace}
\def\cva{{\rm CVA} \xspace}
\def\fva{{\rm FVA} \xspace}

\def\ca{{\rm CA} \xspace}
\def\loss{L}
\def\ca{{\rm CA}\xspace} 

\def\Q{{\mathbb P}}\def\Q{{\mathbb Q}}\def\Q{{\mathbb Q}^*}

\counterwithin{figure}{section}
\counterwithin{table}{section}

\usepackage{wrapfig}
\usepackage{multicol}
\usepackage{comment}

\newcolumntype{x}[1]{>{\centering\arraybackslash\hspace{0pt}}p{#1}}

\newtheorem{theorem}{Theorem}
[section]
\newtheorem{lem}{Lemma}
[section]
\newtheorem{pro}{Proposition}
[section]
\newtheorem{cor}{Corollary}
[section]
\newtheorem{conj}{Conjecture}
[section]

\newtheorem{rem}{Remark}
[section]
\newtheorem{com}{Comments}[section]
\newtheorem{ex}{Example}[section]
\newtheorem{nota}{Notation}[section]
\newtheorem{defi}{Definition}[section]
\newtheorem{hyp}{Assumption}[section]

\newcommand{\bt}{\begin{theorem}}\newcommand{\et}{\end{theorem}} 
\newcommand{\bl}{\begin{lem}}\newcommand{\el}{\end{lem}}
\newcommand{\bp}{\begin{pro}}\newcommand{\ep}{\end{pro}}
\newcommand{\bcor}{\begin{cor}}\newcommand{\ecor}{\end{cor}}
\newcommand{\bconj}{\begin{conj}}\newcommand{\econj}{\end{conj}}

\newcommand{\bd}{\begin{defi} \rm}
\newcommand{\ed}{\finproof\end{defi} }
\newcommand{\eds}{\end{defi} }
\newcommand{\brem }{\begin{rem} 
}\newcommand{\erem }{
\end{rem}}
\newcommand{\bcom}{\begin{com} \rm }\newcommand{\ecom }{\end{com}}
\newcommand{\brems }{\begin{rem} %\rm 
}
\newcommand{\erems }{\end{rem}}
\newcommand{\bex}{\begin{ex} \rm}
\newcommand{\eex}{\finproof\end{ex}}
\newcommand{\bno}{\begin{nota} \rm}
\newcommand{\eno}{\finproof\end{nota}}\newcommand{\enos}{\end{nota}}
\newcommand{\eexs}{\end{ex} }
\newcommand{\bhyp}{\begin{hyp} \rm }\newcommand{\ehyp}{\end{hyp}}
 
\newcommand{\bP}{\begin{proof}}
\newcommand{\eP}{\end{proof}}
\def\proof{\noindent {\it Proof. $\, $}}\def\proof{\noindent{\it\textbf{Proof}. $\, $}}
\def\finproof {\hfill $\Box$ \vskip 5 pt }
\def\finproof{\rule{4pt}{6pt}}
\def\Nom{{\rm  nom}}
\def\VM{{\rm VM}}

\def\VMo{\overline{\rm VM}}

\def\mtm{P\xspace}\def\mtm{{\rm MtM} \xspace}

\def\VaR{\mathbb{V}\mathrm{a}\mathbb{R}'}

\def\SLOIM{{\rm SLOIM}}
\def\theaim{a_{im}}

\usepackage{footnote}
\makesavenoteenv{tabular}
\makesavenoteenv{table}
\usepackage{booktabs,bigstrut}
\usepackage{algorithm}
\usepackage[english]{babel}
\usepackage{amsmath,amssymb,xcolor}

\def\A{\mathfrak{A}}
\def\CM{{\rm CM}\xspace}

\def\kva{{\rm KVA} \xspace}

\def\cva{{\rm CVA} \xspace}
\def\bcva{{\rm BCVA} \xspace}
\def\bmva{{\rm BMVA} \xspace}

\def\fva{{\rm FVA} \xspace}
\def\ca{{\rm CA} \xspace}
\def\loss{L}\def\loss{\g{\cC+\cF}}\def\loss{\cC+\cF}
\def\ca{{\rm CA}\xspace}

\def\cPs{\mathcal{P}^\circ}

\def\mtm{P\xspace}\def\mtm{{\rm MtM} \xspace}

\def\ES{\mathbb{ES}\xspace}
\def\EC{{\rm EC}\xspace}

\def\CMVA{{\rm CMVA} \xspace}
\def\MVA{{\rm MVA} \xspace}

\def\ES{\mathbb{ES}'\xspace}\def\ES{\mathbb{ES}\xspace}

\def\Ep{\mathbb{E}'\xspace}\def\Ep{\mathbb{E}\xspace}
\def\Qp{\mathbb{Q}'\xspace}\def\Qp{\mathbb{Q}\xspace}
\def\Qs{\mathbb{R}\xspace} \def\Qs{\mathbb{Q}^\star\xspace}
\def\VaR{\mathbb{V}\mathrm{a}\mathbb{R}'\xspace}\def\VaR{\mathbb{V}\mathrm{a}\mathbb{R}\xspace}
\def\Es{\mathbb{E}^\star\xspace}

\def\CVA{\cva} 
\def\UCVA{\cva}
 
\def\UFVA{\fva}
\def\CA{\ca} 
\def\KVA{\kva}
\def\cC{\mathcal{C}}
\def\EC{{\rm EC}}
\def\FTP{{\rm FTP}}

\def\bal{\begin{aligned}}
\def\eal{\end{aligned}}
\def\bll#1{
\beqa\label{#1}\bal
}\def\lel{\eal\eeqa}

\newcommand{\DF}{{\rm{DF}}}

\def\cF{\mathcal{F}}

\def\PIM{{\rm IM}}
\def\DF{{\rm DFC}}\def\DF{{\rm DF}}
\def\Ss{\bar{J}} \def\Ss{J}
\def\So{\bar{J}} \def\So{J}
\def\cPo{{\cP}} 
\def\VMo{{\rm VM}}\def\VMo{\mtm} 
\def\PIMo{{\rm IM}}
\def\RIMo{\PIMo}
\def\cPs{{\cP}}\def\cPs{\overline{\cP}} 
\def\VMs{\overline{\rm VM}}\def\VMs{\overline{\mtm}} 
\def\PIMs{\overline{\rm IM}}
\def\IMs{\PIMs}

\def\mtm{{\rm MtM}}
\def\IMo{\PIMo}
\def\CCVA{{\rm CCVA}}
\def\cL{\mathcal{L}\xspace}

\renewcommand{\b}[1]{{\color{blue}#1}}\renewcommand{\b}[1]{#1}
 
\def\tilde{}

\def\then{N} \def\then{n}

\newcommand{\eqdef}{\mathrel{\mathop:}=}

\def\cW{\mathcal{W}}

\def\zeta{\widetilde{\gamma}}

\def\novations{portings\xspace}
\def\novation{porting\xspace}
\def\Novation{Porting\xspace}

\hypersetup{
  colorlinks   = false, %Colours links instead of ugly boxes
  linkbordercolor = {white},
  citecolor   = {black} %Colour of citations
}
\makeatletter
\renewcommand\@makefnmark{\hbox{\@textsuperscript{\normalfont\color{blue}\@thefnmark}}}
\renewcommand\@makefntext[1]{%
  \parindent 1em\noindent
            \hb@xt@1.8em{%
                \hss\@textsuperscript{\normalfont\@thefnmark}}#1}
\makeatother
\begin{document}
\title{Derivatives Risks as Costs in a One-Period Network Model}

\author{{Dorinel Bastide\textsuperscript{a},
St\'ephane Cr\'epey\textsuperscript{b}}, Samuel Drapeau\textsuperscript{c}, Mekonnen Tadese\textsuperscript{d}}   

\date{{\small{}{}{}{}This version: \today}}
\maketitle

\begin{abstract}
We present a one-period XVA model encompassing bilateral and centrally cleared trading in a unified framework with explicit formulas for most quantities at hand. We illustrate possible uses of this framework for running stress test exercises on a financial network from a clearing member's perspective or for
optimizing the \novation of the portfolio of a defaulted clearing member.
\end{abstract}

{\let\thefootnote\relax\footnotetext{\textsuperscript{a} \textit{BNP Paribas Stress Testing Methodologies \& Models. This article represents the opinions of the author, and it is not meant to represent the position or opinions of BNP Paribas or its members.} dorinel.2.bastide@bnpparibas.com  \textbf{(corresponding author).}}}

{\let\thefootnote\relax\footnotetext{\textsuperscript{b} \textit{Laboratoire de Probabilités, Statistique et Modélisation (LPSM), Sorbonne Université et Université
de Paris, CNRS UMR 8001.} stephane.crepey@lpsm.paris.}}

{\let\thefootnote\relax\footnotetext{\textsuperscript{c} \textit{Shanghai Jiao Tong University, Shanghai, China.} sdrapeau@saif.sjtu.edu.cn.}}

{\let\thefootnote\relax\footnotetext{\textsuperscript{d} \textit{Woldia University,
Mathematics Department, Ethiopia.} mekonnenta@wldu.edu.et.}}

{\let\thefootnote\relax\footnotetext{\textit{Acknowledgments:} We thank
Paul Besson, Head of Quantitative research, Euronext, and
Mohamed Selmi, Head of Market Risk - LCH SA, for useful discussions. 
This work benefited from the support of the grant \textit{When Credit Meets Liquidity: The Clearing Member Default Resolution Issue}, under the aegis of the Europlace Institute of Finance, France. 
The research of S. Crépey benefited from the support of the \textit{Chair Stress Test, RISK Management and Financial Steering}, led by the French \'Ecole
polytechnique and its Foundation and sponsored by BNP Paribas.}}

\section{Introduction\label{s:intro}}

In the wake of the 2008--09 global financial crisis, clearing through central counterparties (CCPs) has become mandatory for standardized derivatives, other ones remaining under bilateral setup with higher capital requirements.

One role of the CCPs\footnote{See  \cite{Gregory14} and \citeN{gregory2015xva} for general CCP and XVA references,
as well as \citeN{MenkveldVuillemey2021} for a recent CCP survey.}
is to provide to their clearing members fully collateralized hedges of their market risk with their clients.  
But this comes at a cost to the clearing members, which pass it to their corporate clients in the form of XVA (cross-valuation adjustments) add-ons.   
Bearing in mind that
the risks of a hedge are, by definition, of the same magnitudes 
as the ones of the originating position and that standardized derivatives usable as hedging assets have to be traded through CCPs,
the XVA footprint of not only bilateral but also centrally cleared trading
is significant and should be analyzed in detail, which is the topic of this paper. 

More precisely, the trades of a clearing member bank with a CCP are partitioned between proprietary trades, which are in effect hedges of the bilateral trading exposure of the bank, and back-to-back hedges of so-called cleared client trades, through which non-member clients gain access to the clearing services of a CCP. 
\citeN{AAC20} focus on the XVA analysis of a bank only acting as a clearing member of one CCP, without proprietary trading.
The present paper provides  
an integrated XVA analysis in the realistic situation of a bank dealing with many clients and CCPs, through both proprietary (also dubbed house) accounts and client accounts.

For the sake of tractability, this is achieved in a stylized one-period setup, fine-tuned to applications including risk assessment in the context of stress test exercises\footnote{as required by Article 302 of the CRR document \citeN{EU2013}.} or optimizing the \novation of the portfolio of defaulted clearing members.

\begin{figure}[ht]
\begin{center}
\includegraphics[width=1\textwidth]
{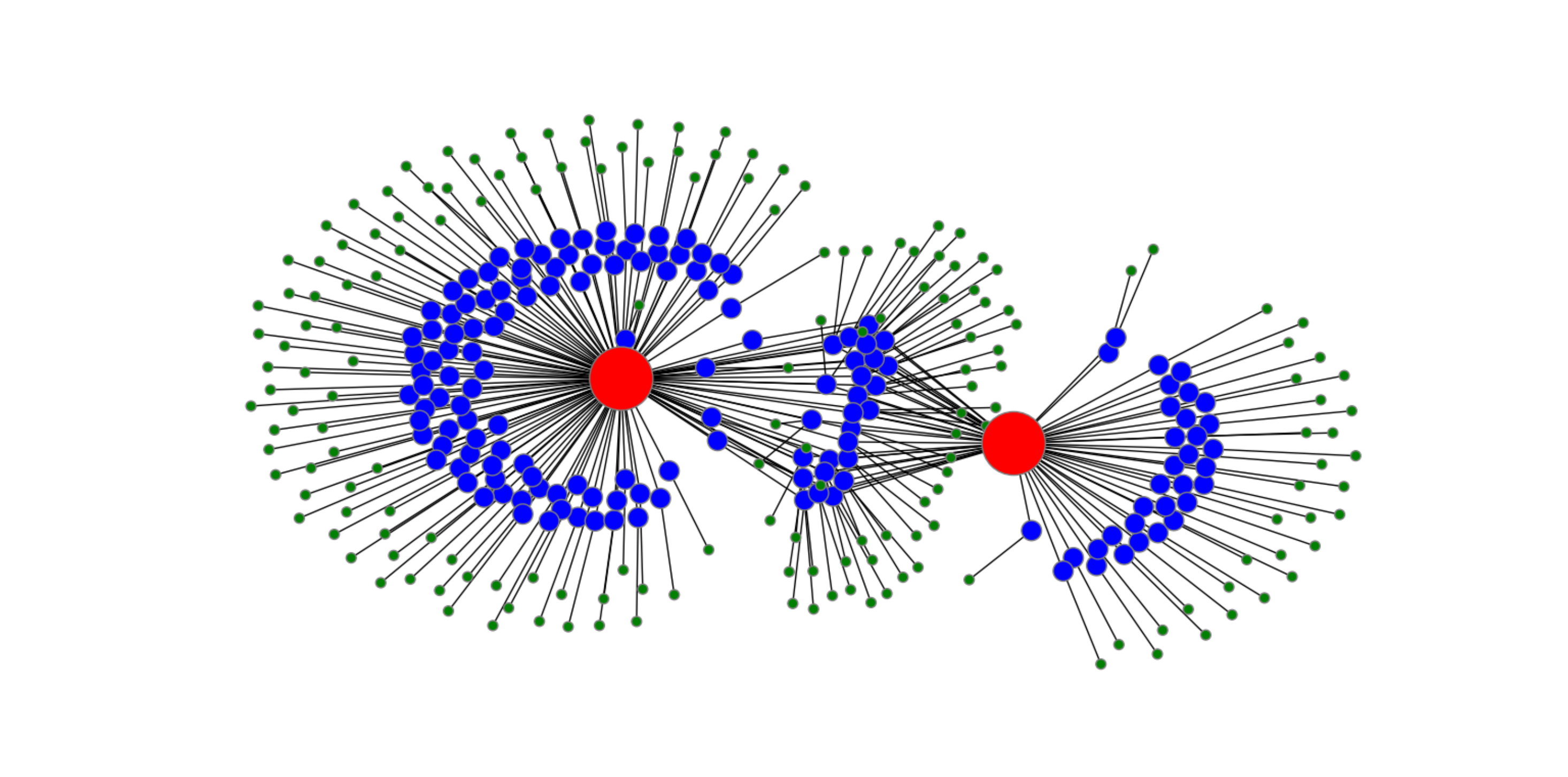}
\end{center}
\caption{Network consisting of two CCPs (in red), 123 members for CCP1 seen on the left hand side, and 56 members for CCP2 on the right hand side, with 24 common members displayed as the group of members in the middle of the two CCPs (155 members in total, in blue), and with 179 cleared clients (in green).}
\label{fig:Network2CCPsUC}
\end{figure}

{The first type of application is motivated by the default in 2020 of Ronin Capital, a broker/dealer firm that had clearing exposures on both CCP services Fixed Income Clearing Corporation (FICC) GSD\footnote{Government Securities Division.} segment (123 members) and CME Futures (56 members of which 24 common with FICC GSD). If all members are assumed to be only exposed to these CCPs and their cleared clients, we can illustrate these relationship by the network depicted in \hyperref[fig:Network2CCPsUC]{Figure \ref{fig:Network2CCPsUC}}. Any common member on those two CCPs needs to ensure conservative risk assessment that can be achieved in the proposed framework by accounting for common memberships on the two CCPs. If such common memberships are ignored, they can lead to lower loss estimates giving wrong risk view on potential losses.}

{The second type of application is an illustration of the results of defaulted portfolio \novation as it has been the case for the trader Einer Aas on NASDAQ OMX\footnote{\textbf{O}ptions\textbf{m}äklarna/Helsinki Stock E\textbf{x}change} that has defaulted on 2018 with loss spill-over effect on surviving members.} 

Section \ref{s:setup} sets the stage.
Section \ref{s:xvaa} develops the corresponding XVA analysis.
Section \ref{s:topract} develops two applications in the above veins.
Section \ref{sec:concl} concludes.

\section{General Setup\label{s:setup}}
 
We consider a finite set of market participants, also susceptible to serve as clearing members of CCPs. 
Derivative transactions can then be concluded between two individual participants, or between a set of participants\footnote{two or more, in practice from a few units to a few hundreds.}, pooled in the form of a CCP, and a clearing member of this CCP.

\begin{figure}
\begin{center}
\begin{tikzpicture}[
roundnodeC/.style={circle, draw=white, fill=white, thick, minimum size=10mm},
roundnodeB/.style={circle, draw=green!60, fill=green!5, very thick, minimum size=10mm},
roundnodeCH/.style={circle, draw=black, fill=red!5, very thick, minimum size=25mm},
roundnodeJ/.style={circle, dashed, draw=black, fill=white, thick, minimum size=80mm},
scale=0.7, transform shape]

%Nodes
\node at (0,0) [roundnodeJ]  (theJoint)    {};

\draw[thick, dashed, green] (6,3.6) arc (-45:-160:6);
\node at (-3.2,4) [rotate=-47, green] {CCP};
\node at (-3.8,2) [rotate=60, black] {CCP};

\draw[thick, dashed, violet] (5,-5.1) arc (45:145:6);
\node at (-3.2,-4.5) [rotate=42, violet] {CCP};

\node at (0,0) [roundnodeCH]  (theCCP)    {};

\node at (3.5,2) [roundnodeB]   (bank3)     {CM};
\node at (6, 4) [roundnodeC]   (c3c)     {};
\node at (5.8, 4.5) [roundnodeC]   (c3b)     {};
\node at (5.5, 5) [roundnodeC]   (c3a)     {};
\node at (6.5, 3) [roundnodeC]   (c3d)     {};
\node at (6.4, 2.5) [roundnodeC]   (c3e)     {};
\node at (6.3, 2) [roundnodeC]   (c3f)     {};

\node at (6, -4) [roundnodeC]   (c4c)     {};
\node at (5.8, -4.5) [roundnodeC]   (c4b)     {};
\node at (5.5, -5) [roundnodeC]   (c4a)     {};
\node at (6.5, -3) [roundnodeC]   (c4d)     {};
\node at (6.4, -2.5) [roundnodeC]   (c4e)     {};
\node at (6.3, -2) [roundnodeC]   (c4f)     {};

\node at (3.8, -5.6) [roundnodeC]   (c5c)     {};
\node at (3.5, -5.8) [roundnodeC]   (c5b)     {};
\node at (3.2, -6) [roundnodeC]   (c5a)     {};
\node at (2.7, -6) [roundnodeC]   (c5d)     {};
\node at (2.4, -5.8) [roundnodeC]   (c5e)     {};
\node at (2.1, -5.6) [roundnodeC]   (c5f)     {};

\node at (-4,0) [roundnodeB]   (bank1)     {CM};
\node at (-7, 0.5) [roundnodeC]   (c1c)     {};
\node at (-6.8, 1) [roundnodeC]   (c1b)     {};
\node at (-6.5, 1.5) [roundnodeC]   (c1a)     {};
\node at (-7, -0.5) [roundnodeC]   (c1f)     {};
\node at (-6.8, -1) [roundnodeC]   (c1e)     {};
\node at (-6.5, -1.5) [roundnodeC]   (c1d)     {};
\draw[thick, blue, ->] (c1a.east) -- (bank1.110);
\draw[thick, blue, ->] (c1b.east) -- (bank1.130);
\draw[thick, blue, ->] (c1c.east) -- (bank1.150);
\draw[thick, orange, ->] (c1d.east) -- (bank1.-110);
\draw[thick, orange, ->] (c1e.east) -- (bank1.-130);
\draw[thick, orange, ->] (c1f.east) -- (bank1.-150);

\node at (4,0) [roundnodeB]   (bank2)     {CM};
\node at (-4,0) [roundnodeB]   (bank1)     {CM};
\node at (7, 0.5) [roundnodeC]   (c2c)     {};
\node at (6.8, 1) [roundnodeC]   (c2b)     {};
\node at (6.5, 1.5) [roundnodeC]   (c2a)     {};
\node at (7, -0.5) [roundnodeC]   (c2d)     {};
\node at (6.8, -1) [roundnodeC]   (c2e)     {};
\node at (6.5, -1.5) [roundnodeC]   (c2f)     {};
\draw[thick, blue, ->] (c2a.west) -- (bank2.70);
\draw[thick, blue, ->] (c2b.west) -- (bank2.50);
\draw[thick, blue, ->] (c2c.west) -- (bank2.30);
\draw[thick, orange, ->] (c2d.west) -- (bank2.-30);
\draw[thick, orange, ->] (c2e.west) -- (bank2.-50);
\draw[thick, orange, ->] (c2f.west) -- (bank2.-70);

\draw[thick, blue, ->] (c3a.west) -- (bank3.90);
\draw[thick, blue, ->] (c3b.west) -- (bank3.70);
\draw[thick, blue, ->] (c3c.west) -- (bank3.50);
\draw[thick, orange, ->] (c3d.west) -- (bank3.10);
\draw[thick, orange, ->] (c3e.west) -- (bank3.-10);
\draw[thick, orange, ->] (c3f.west) -- (bank3.-30);

\node at (3.5,-2) [roundnodeB]   (bank4)     {CM};
\draw[thick, orange, ->] (c4a.west) -- (bank4.-90);
\draw[thick, orange, ->] (c4b.west) -- (bank4.-70);
\draw[thick, orange, ->] (c4c.west) -- (bank4.-50);
\draw[thick, blue, ->] (c4d.west) -- (bank4.-10);
\draw[thick, blue, ->] (c4e.west) -- (bank4.10);
\draw[thick, blue, ->] (c4f.west) -- (bank4.30);

\node at (2,-3.5) [roundnodeB]   (bank5)     {CM};
\draw[thick, blue, ->] (c5c.west) -- (bank5.-50);
\draw[thick, blue, ->] (c5b.west) -- (bank5.-70);
\draw[thick, blue, ->] (c5a.west) -- (bank5.-80);
\draw[thick, orange, ->] (c5d.west) -- (bank5.-100);
\draw[thick, orange, ->] (c5e.west) -- (bank5.-110);
\draw[thick, orange, ->] (c5f.west) -- (bank5.-120);

%Lines
\draw[thick, orange, ->] (bank1.-10) -- (theCCP.184) node[below, midway]{$\overline{\mathcal{P}}$};
\draw[thick, blue, ->] (bank1.10) -- (theCCP.176) node[above, midway]{$\mathcal{P}$};
\draw[thick, blue, ->] (bank2.170) -- (theCCP.4);
\draw[thick, orange, ->] (bank2.190) -- (theCCP.-4);
\draw[thick, blue, ->] (bank3.190) -- (theCCP.43);
\draw[thick, orange, ->] (bank3.210) -- (theCCP.35);
\draw[thick, blue, ->] (bank4.150) -- (theCCP.-35);
\draw[thick, orange, ->] (bank4.170) -- (theCCP.-43);
\draw[thick, blue, ->] (bank5.100) -- (theCCP.-67);
\draw[thick, orange, ->] (bank5.120) -- (theCCP.-75);

\end{tikzpicture}
\end{center}
\caption{Promised cash flows between market participants. The reference clearing member bank is on the left.}
\label{fig:CFs}
\end{figure}
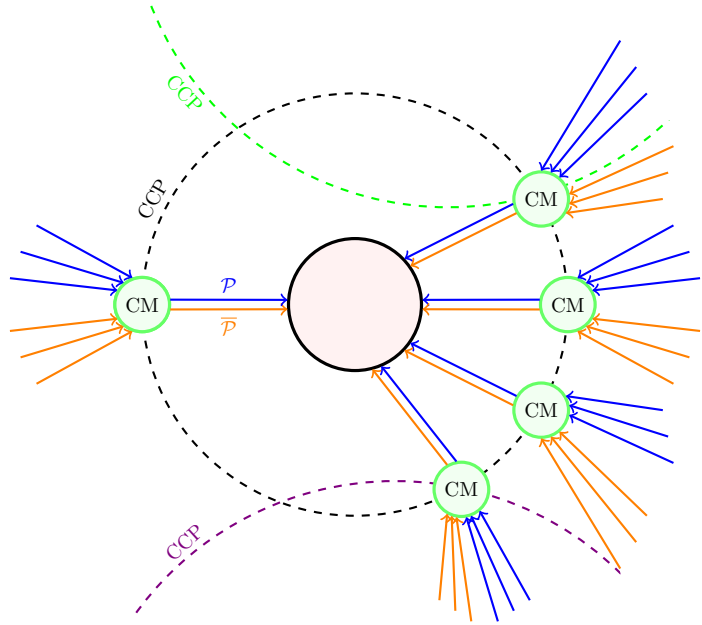

CCPs are typically siloed into different services, each devoted to a specific class of derivatives. 
We first consider a setup with a single CCP service\footnote{the extension to several CCPs is done in \hyperref[ext:ccps]{Section \ref{ext:ccps}}.}: 
see Figure \ref{fig:CFs}, where $\cP$ and $\cPs$ represent the
contractual cash flows from cleared and bilateral clients to a reference clearing member, dubbed the bank hereafter, hence promised, in successive turns, from the bank to the CCP, from the CCP to other clearing members, and from the latter to their own clients.
As a consequence, the CCP is flat in terms of market risk, as is also each of the clearing members.
 
\subsection{Defaults Settlement Rule}

As reasserted in the wake of the 2008--09 global financial crisis by the Volcker rule, a dealer bank
% is not supposed to do proprietary trading, i.e.~it 
 should be hedged as much as possible, at least in terms of market risk\footnote{cf.~paragraph number 1851 in section 619 from \citeN{DF2010}.}. Jump-to-default risk, on the other hand, is hardly hedgeable in practice. Instead it is mitigated through netting and collateralization.
Namely, designated netting sets of transactions between two given counterparties (two individual participants or a participant and the CCP) are jointly
collateralized, i.e.~guaranteed against the default of one or the other party. The collateral (or guarantee) comprises a variation margin, which tracks the mark-to-market (counterparty-risk-free value) of the netting set between the two parties, and nonnegative amounts of initial margin posted by each party to the other, which provide a defense against the risk of slippage of the value of the netting set away from its (frozen) variation margin during its liquidation period.
In the case of transactions with a CCP, there is an additional layer of {collateral} in the form of the (funded) default fund contributions of the clearing members, which is meant as a defense against extreme and systemic risk.
For each participant, variation margin is
rehypothecable and fungible across all its netting sets. Initial margin is segregated at the netting set level.
default fund contributions are segregated at the clearing member level.
 
The general rule regarding the settlement of contracts of a defaulted netting set is that:
\bhyp \label{h:rule}	 If a counterparty in default is indebted toward the other beyond its posted margin, then this debt is
only reimbursed at the level of this posted margin (assuming zero recovery rate of the defaulted  party for simplicity in this paper);
otherwise the debt between the two parties is fully settled.

Here debt is understood on a counterparty-risk-free basis. 
\ehyp

\noindent
This rule also applies to a netting set of transactions between a clearing member and a CCP. However, in our stylized setup, 
a CCP is nothing but the collection of its clearing members. Our CCP has no resources of its own (in particular, it cannot post any default fund contribution, or ``skin-in-the-game''\footnote{such additional protection layer, though quite common in practice, is of marginal magnitude compared to the other protection layers. By omitting skin-in-the-game component, the obtained results are conservative in terms of risk management and the various formulations are simplified.}). 
As long as it is non-default, i.e.~as long as at least one of its clearing members is non-default, our CCP can only handle the losses triggered by the defaults of some of its clearing members by redirecting these losses on the surviving ones. This participation of the surviving members to the losses triggered by the defaults of the other members
corresponds in our framework to the usage by the CCP of their default fund contributions, both funded (as already introduced above) and unfunded.   
As will be detailed in equations below, the funded default fund contributions are used in priority for covering losses triggered by the defaults of clearing members over their margins.
The unfunded default fund contributions correspond to additional refills that can be required by the CCP, often up to some cap in principle, without bounds in our model, in case the funded default fund contributions of the surviving members are not enough.

\subsection{XVA Framework\label{s:xvaf}} 

Assume that at time 0 all the banking participants, including the reference clearing member bank, in Figure \ref{fig:CFs}, with no prior endowments, enter transactions with their clients and hedge their positions, both bilaterally  between them and through the CCP.
As seen above, the CCP and each bank are flat in terms of market risk.
However, as market participants are assumed to be  defaultable with zero recovery, in order to account for counterparty credit risk and its funding and capital consequences, each banking participant requires
from its corporate clients a pricing rebate (considering conventionally the bank as the ``buyer'') with respect to the mark-to-market (counterparty-risk-free) valuation of the deals. 
The corporate clients of the bank are assumed to absorb the ensuing prices via their corporate business, which is their primary motivation for these deals.
 
A reference probability measure $\Qs$, with corresponding expectation operator denoted by $\Es$, is used for the linear valuation of cash flows, using the risk-free asset as our num\'eraire everywhere. This choice of a num\'eraire simplifies equations by removing all terms related to the (assumed risk-free) remuneration of all cash and collateral accounts. The funding issue is then refocused on the risky funding side of the problem, i.e.~funding costs in what follows really means excess funding costs with respect to a theoretical situation where the bank could equally borrow and lend at the risk-free rate.

More precisely, as suitable for XVA calculations \citep[Remark 2.3]{CrepeyHoskinsonSaadeddine2019}: given a physical probability measure defined on the full model $\sigma$ algebra $ \A$ and equivalent to a reference risk-neutral measure on the financial sub $\sigma$ algebra $\mathfrak{B}$ of $ \A$,
we take $\Qs$ equal to the reference risk-neutral measure on $\mathfrak{B}$ and equal to the physical probability measure   conditionally on $\mathfrak{B}$.

Following the general XVA guidelines of
\citeN{CrepeyHoskinsonSaadeddine2019}, the above-mentioned pricing rebate required by the reference clearing member bank,
dubbed funds transfer price (FTP), comes in two parts: first, the expected counterparty default losses and funding expenditures of the bank, an amount that flows into the bank liabilities and which we refer to as contra-asset valuation (CA); second, a cost of capital risk premium (KVA), which instead is loss-absorbing\footnote{hence, not a liability.} and is also used by the management of the bank as retained earnings for remunerating the shareholders of the bank for their capital at risk within the bank.
All in one, the bank buys the deals from its clients at the (aggregated) price $(\mtm-{\rm FTP})$, where $\mtm$ is their counterparty-risk-free value and
\beql{e:bsheet-pedago}
{\rm FTP}&= \underbrace{\CA }_{\mbox{Expected costs}} 
+ \underbrace{\rm KVA}_{\mbox{Risk premium}}.
\eeql

\bhyp \label{h:CAsource}
At time 0 the amounts $\CA$ and $\KVA$ sourced from the corporate clients of the bank are deposited on reserve capital and capital at risk accounts of the bank.
\ehyp

Let EC denote an economic capital of the bank corresponding to the minimum level of capital at risk that the bank should hold from a regulatory (i.e.~solvency) perspective. If $\KVA<\EC$, then the bank shareholders need to provide the missing amount $(\EC-\KVA)$ of capital at risk, so that the actual level of capital at risk of the bank is
$$\max(\EC,\KVA),$$ 
while \textit{shareholder} capital at risk reduces to
\beql{e:scr}\max(\EC,\KVA)-\KVA=(\EC-\KVA)^+.\eeql

\section{Theoretical XVA Analysis\label{s:xvaa}} 

In this section we detail each term in the equations above, in the realistic setup of a bank involved into an arbitrary combination of bilateral and centrally cleared portfolios, in a tractable one-period setup with period length $T$.
In the one-period XVA model of \citeN[Section 3]{CrepeyHoskinsonSaadeddine2019}, there were no CCPs and the bank was assumed to have access to a ``fully collateralized back-to-back hedge of its market risk", ensuring by definition and for free to the bank a cash-flow $(\cP-\mtm )$ at time 1, irrespective of the default status of the bank and its client. There, $\cP$ denoted the 
contractual cash flows from the (assumed unique) client to the bank and MtM was the corresponding counterparty-risk-free value.
In the present paper we reveal the mechanism of such a ``fully collateralized hedge of the market risk'' of the bank, which can be achieved through central clearing, but at a certain cost that we analyze.\smallskip

\subsection{Cash Flows\label{s:cfs}}

We use the terms client for cleared clients and counterparty for bilateral counterparties.

Given disjoint sets of indices $I\ni 0$, $C$, and $B$  for the clearing members (including the reference bank labeled by 0) and for the respective cleared and bilateral netting sets of the bank with its individual clients and counterparties,
We denote by:
\begin{itemize} 
\item  $\S_0$, shortened as $J$, and $\S_i,$ $i\in I\setminus\{0\}$,
the survival indicator random variables of the bank and of the other clearing members at time 1;  
$\gamma=\Qs(\S=0),$ the default probability of the bank; 
\item $\cJ=\max_i \S_i$, the  survival indicator random variable of the CCP (i.e.~of at least one clearing member),
\item  $\cP_i $, $\mtm_i 
 =\Es \cP_i 
$, and $ \PIM_i$, $i\in I$, the contractual cash flows, variation margin, and  initial margin from the clearing member $i$ to the CCP corresponding to the  
cleared clients account of the member $i$; 
\item  $\cPs_i $, $\VMs_i= \Es \cPs_i  
$, and $ \PIMs_i $,
, $i\in I$, the contractual cash flows, variation margin, and initial margin from the clearing member $i$ to the CCP  corresponding to the 
house account 
of the clearing member $i$;   
\item  $ \DF_i$, $i\in I$, the default fund contribution posted by the clearing member $i$ to the CCP;   
\item  $\S_b$, $b\in B$,
the survival indicator random variable of the
%client
counterparty of the bilateral netting set $b$ of the reference bank; $\cP_b $, $\VM_b $, and $ \PIM_b$, the  associated contractual cash flows, variation margin, and initial margin from the corresponding counterparty to the bank; and  $\overline{\PIM}_b$, the initial margin from the bank to the counterparty;
\item  $\S_c$, $c\in C$,
the survival indicator random variable of the client of the cleared trading netting set $c$ of the bank, and $\cP_c $, $\mtm_c
 =\Es \cP_c  
$\footnote{reflecting the fact that members of CCPs are fully collateralized.}, and $ \PIM_c$, the associated contractual cash flows, variation margin, and initial margin  from the corresponding client to the bank\footnote{note that a bank does not post any initial margin on its cleared client netting sets.};  

\item $\cL $, the loss of the CCP, i.e.~the loss triggered by the defaults of its clearing members beyond their posted collateral\footnote{variation margin, initial margin, and (funded) default fund contributions.}, 
which is borne by the surviving members (if any);
\item  $\mu=J\mu$, the proportion of these losses allocated to the reference clearing member bank.
\end{itemize}

\bhyp\label{h:contract}  $\sum_i ( \cP_i +   \cPs_i)=0$ (the CCP is flat in terms of market risk),
$\sum_c  \cP_c  =\cP_0  $ (by definition of cleared trades and of their mirroring trades),
 and
$  \sum_b  \cP_b= \cPs_0  $ (the reference bank is flat in terms of market risk).\ehyp

Assumption \ref{h:contract} yields the clearing conditions regarding the contractually promised cash flows, which applies to each banking participant (written there for the reference bank) and to the CCP.
Assumption \ref{h:rule} is monitoring the default cash flows. We need one more condition, regarding the funding side of the problem:

\bhyp \label{h:funding} The bank can use the amounts $\CA$ and $\max(\EC,\KVA)$ on its reserve capital and capital at risk accounts for its variation margin borrowing purposes. 
Funds needed beyond $\CA+\max(\EC,\KVA)$ for variation margin posting purposes are borrowed by the bank at \b{its credit spread $\gamma$} above OIS.
The initial margin and default fund contributions, instead, must be
borrowed entirely by the bank, but this can be achieved at some blended funding spread $\widetilde{\gamma}\le\gamma$.
\ehyp
The rationale for funding variation margin but not initial margin from $\CA+\max(\EC,\kva)$ is set out before Equation (15) in \citeN{AlbaneseCaenazzoCrepey17b}. The motivation for the assumption $\widetilde{\gamma}\le\gamma$ is provided in \citeN[Section 5]{AAC20}, along with numerical experiments suggesting that $\widetilde{\gamma}$ can be several times lower than $\gamma$.

\bl\label{lem:vmfun}
The borrowing needs of the bank for reusable and segregated collateral amount to, respectively,
\beql{eq:cF}& \big(\sum_b (\mtm_b-{\rm VM}_b) -\CA-\max(\EC,\KVA) \big)^+ \\
&  {\IM  +\PIMs+\DF  +  \sum_b \IM_b}.
\eeql
\el
\proof
On the bilateral trades of the bank and their hedges, the Treasury of the bank receives $\sum_b {\rm VM}_b$ of variation margin from the counterparties and has to post an aggregated amount $\sum_b \mtm_b$ of variation margin. 
The assumption stated before the lemma then leads to \qr{eq:cF}.
\finproof
 
\bl\label{l:cfoneper} 
On the bank survival event $\{J=1\}$, the counterparty default losses $\cC$ and the funding  expenses $\cF$ of the bank are given by
\beql{e:recapprelccpc}
& \cC=  
\sum_c(1-\So_c)(\cPo_c-\VMo_c-\IMo_c)^+    +\mu\cL   \\&\qqq
+\sum_b  (1-\Ss_b)(\cP_b-{\rm VM}_b-\IM_b )^+,\eeql
where 
 {\beql{e:L} 
\cL  = 
\sum_i (1-J_i)\big((\cP_i -\mtm_i - \PIM_i )^+ + (\cPs_i -\VMs_i - \PIMs_i )^+  -\DF_i\big)^+ 
 ,\eeql} 
and
\beql{e:recapprelccpf}
&\cF=  {\zeta}\big( \IM  + \PIMs + \DF\big)+
 \\&
 \qqq
 + \zeta \sum_b  {  \IMs_b} +   \gamma   \big(\sum_b (\mtm_b-{\rm VM}_b) -\CA-\max(\EC,\kva) \big)^+ .\eeql 
\el  
  
\proof On the CCP survival event $\{\cJ=1\}$, the CCP receives, by Assumption \ref{h:rule},
\beql{e:cfccp}&\sum_i \bigg(  J_i (\cP_i +\cPs_i) + (1-J_i)\Big(
 \cP_i \wedge (\mtm_i + \PIM_i )
+\cPs_i \wedge (\VMs_i + \PIMs_i )+\\&\qqq
\big((\cP_i- (\mtm_i + \PIM_i ))^+ +
(\cPs_i - (\VMs_i + \PIMs_i ))^+ \big) \wedge \DF_i \Big)\bigg).
\eeql 
Using the CCP clearing condition Assumption \ref{h:contract}: 
$$0=\sum_i ( \cP_i +   \cPs_i)=\sum_i\big(  J_i (\cP_i +\cPs_i) + (1-J_i)  ( \cP_i +   \cPs_i) \big)$$  
in Assumption \ref{h:contract},
\eqref{e:cfccp} is equal to
\bel&-\sum_i  (1-J_i) \big((\cP_i -\mtm_i - \PIM_i )^+ + (\cPs_i -\VMs_i - \PIMs_i )^+  -\DF_i\big)^+ =-\cL ,\eel 
which is \eqref{e:L}.

On the bank survival event $\{J=1\}$ ($\subseteq \{\cJ=1\}$), 
by the respective Assumptions \ref{h:rule} and \ref{h:contract}, the bank receives from its clients and counterparties
\beql{e:rec}
    &\sum_{c}\Big(\So_c  \cPo_c  +(1-\So_c )\big(\cPo_c  \wedge (\VMo_c  +\RIMo_c  )\big)\Big)+  \sum_b\Big(\So_b  \cPo_b  +(1-\So_b )\big(\cPo_b  \wedge (\VM_b  +\RIMo_b  )\big)\Big) , \eeql
respectively pays to the CCP
\beql{e:pay}
  &\sum_{c}   \cPo_c + \sum_b    \cPo_b  = 
   \sum_c \big(\So_c  \cPo_c  +(1-\So_c )\ \cPo_c   \big) +\sum_b \big(\So_b  \cPo_b  +(1-\So_b )\ \cPo_b   \big) . \eeql
Subtracting  \eqref{e:rec} from  \eqref{e:pay}, we obtain  
$$ \sum_c (1-\So_c )(\cPo_c   - \VMo_c  -\RIMo_c    )^+ + \sum_b (1-\Ss_b)(\cP_b-{\rm VM}_b-\IM_b )^+ .$$
On top of this comes the participation $\mu\cL $ of the bank to the CCP default losses, which yields \eqref{e:recapprelccpc}. 
Moreover, in view of Lemma \ref{lem:vmfun} and Assumption \ref{h:funding}, the (risky) funding  expenses of the bank are given by \eqref{e:recapprelccpf}.~\finproof  
\subsection{Valuation} 
 
Let $\Ep$ denote the expectation with respect to the bank survival measure $\Qp$ associated with $\Qs$, i.e., for any random variable $\cY$,
\beql{e:survdi}
&\Ep\cY
=(1-\gamma)^{-1}\Es( \S \cY).
\eeql
(expectation of $\cY$ conditional on the survival of the bank).
As easily seen in \citeN[Section 3]{CrepeyHoskinsonSaadeddine2019}:
\bl \label{e:thelem} For any random variable $\cY$ and constant $Y$, we have 
\bel
Y =\Es (J\cY + (1-J)Y)  \Longleftrightarrow Y=\Ep   \cY.
\eel
\el

Under a cost-of-capital XVA approach, the bank charges its future losses to its corporate clients at a CA level 
making $J( \loss-\ca)$, the trading loss of the shareholders of the bank, $\Qs$ centered. In addition, given a target hurdle rate $h$
assumed in $[0,1]$ (and typically of the order of $10\%$),
 the management of the bank ensures to the bank shareholders dividends at the height of $h$ times their capital at risk $(\EC-\mathrm{KVA})^+$ (cf.~\eqref{e:scr}), where we model EC as $ \ES\big(J( \loss-\ca)\big) $, the expected shortfall of the trading loss $\ell=J( \loss-\ca)$
computed under the bank survival measure at a quantile level\footnote{under normal distribution assumptions, such ES at percentile level $99.75\%$ allows reaching similar loss level as with a VaR (quantile) risk metric at the level $99.9\%$. In practice, regulatory and economic capital indeed aims at capturing extreme losses that can occur once every 1000 years, cf.~paragraph 5.1 from \citeN{BCBS05} for the detailed instructions.} of $\alpha=99.75\%$, i.e. under the primal and dual representations of the expected shortfall\footnote{see e.g. \citeN[Equations (2.1) and (5.5)]{follmer2010convex}.}, 
 $\VaR_a(\ell)$ denoting the $\Qp$ value-at-risk (lower quantile) of level $a$ of $\ell$:
 \beql{e:es}
 \begin{array}{lcl}
      \EC & = &  \frac{1}{1-\alpha} \int_{a=\alpha}^1   \VaR_a (\ell)
da \\
      & & \\
      & = & \sup  \left\{ \Ep [{\ell\chi}] \;;\; \chi \mbox{ is measurable, }
0\leq \chi \leq(1-\alpha)^{-1}, \mbox{ and } \Ep [\chi] =1 \right\},
 \end{array}
\eeql 
 which for atomless $\ell$ also coincides\footnote{see Corollary 5.3 and representation thanks to expression (3.7) from \citeN{AT2002}.} with $ \Ep [\ell | \ell\ge \alpha] .$
Note that, in view of the dual representation, an expected shortfall of a centered random variable is nonnegative.
 
Accordingly (as detailed after the definition):

\bd\label{d:ccr} We let
\beql{e:ccvadi} 
\CA =\; &
\underbrace{\CVA}_{\bcva+\CCVA}+\underbrace{\MVA}_{\bmva +\CMVA}+\fva, \mbox{  where }\\
   & \CCVA=\Es\Big(J\big(\sum_c (1-\So_c )(\cPo_c -\VMo_c -\IMo_c )^+  
  +\mu\cL  \big) +(1-\S)\CCVA \Big) ,\\
 &\CMVA=\Es\Big( J\zeta\big( \IM+ \PIMs + \DF \big) +(1-\S)\CMVA \Big) ,\\
  &\bcva=\Es\big( J\sum_b  (1-\Ss_b)(\cP_b-{\rm VM}_b-\IM_b )^+     +(1-\S)\bcva \big) ,\\
   &\bmva=\Es\big( J \zeta\sum_b   \IMs_b  +(1-\S)\bmva \big) ,\\
&\fva=\Es\Big( J\gamma\big(\sum_b (\mtm_b-{\rm VM}_b) -\CA-\max(\EC,\kva) \big)^+  +(1-\S)\fva\Big), \\
\kva=\;&\Es\big( J  h (\EC-\mathrm{KVA})^+ +(1-J)\kva \big), \mbox{  where }\EC= \ES\big(J( \loss-\ca)\big).
\eeql    
\eds

Hence in view of \eqref{e:recapprelccpc} and \eqref{e:recapprelccpf}:
\beql{e:casynth}
\ca= \Es\Big(J\big(\cC+\cF  \big) +(1-\S)\ca \Big)  ,
\eeql
i.e. $  \Es\Big(J\big(\cC+\cF  -\ca )\Big)=0 $, as desired\footnote{see after Lemma ~\ref{e:thelem}.}.
The terminal cash flows of the form $(1-J)\times \cdots$ in \eqref{e:ccvadi} or \eqref{e:casynth} are thus consistent with the desired shareholder centric perspective. They can also be interpreted as the amounts of reserve capital and risk margin lost by the bank shareholders, as their property is transferred to the liquidator of the bank, if the bank defaults.

Due to these terminal cash flows, the above definition is in fact a fix-point system of equations. 
The split of the underlying CA equation \eqref{e:casynth} into  the collection of equations \eqref{e:ccvadi} is motivated by both interpretation and numerical considerations. From an interpretation viewpoint, it is useful to provide the more granular view on the costs of the bank provided by the split of the global CA amount between, on the one hand, bilateral and centrally cleared trading default risk components BCVA and CCVA and, on the other hand, bilateral and centrally cleared trading funding risk components BMVA and CMVA for segregated initial margin, whereas  the FVA cost of funding variation margin is holistic in nature (can only be apprehended at the level of the bank balance-sheet as a whole), via the feedback impact of $\CA+\max(\EC,\kva) $ into the FVA.
From a numerical viewpoint, the collection \eqref{e:ccvadi} of smaller problems may be easier to address than the global equation \eqref{e:casynth}.  Each of the smaller problems can also be handled by a dedicated desk of the bank, namely
the CVA desk, for the  BCVA and CCVA, and the Treasury of the bank, for the BMVA, CMVA and the FVA.

Passing in the above equations to the bank survival measure $\Q$ based on Lemma \ref{e:thelem} shows that the corresponding fixed point problem is in fact well-posed and yields explicit formulas for all the quantities at hand.

\bp\label{p:ccpr}
We have 
\beql{L:discrs} 
&  \CCVA 
=  \Ep \big[ \sum_c (1- \So_c )(\cPo_c -\VMo_c -\IMo_c )^+
  +\mu\cL \big], \mbox{where}\\ &\qqq\cL = \displaystyle\sum_i  (1-\S_i ) \big(( \mathcal{P}_i -\mtm_i -\IM_i 
 )^+ +( \cPs_i -\VMs_i -\PIMs_i 
 )^+ -\DF_i \big)^+,\\
&\CMVA
= \zeta\big(\IM +\PIMs + \DF \big) ,\\ 
  &\bcva=\Ep \big( \sum_b  (1-\Ss_b)(\cP_b-{\rm VM}_b-\IM_b )^+   \big) ,\\
   &\bmva=\zeta\sum_b  \IMs_b 
 \\&\EC=\ES(J(  \cC-\cva))\ge 0, \mbox{ where }\\
   &\qqq  J(  \cC-\cva) = J\Big(\sum_c(1-\So_c)(\cPo_c-\VMo_c-\IMo_c)^+    +\mu\cL  -  \CCVA \\&\qqq\qqq+
\sum_b (1-\Ss_b)(\cP_b-{\rm VM}_b-\IM_b )^+  -\bcva\Big),  \\
&\fva=\frac{\gamma}{1+\gamma}   \big(\sum_b (\mtm_b-{\rm VM}_b) -
(\CCVA+\CMVA+\bcva +\bmva)- \EC  \big)^+  , \\
 & \kva=\frac{h}{1+h} \EC.
\eeql 
All the above XVA numbers are nonnegative.
\ep
\proof By the result recalled after \eqref{e:es},
EC is nonnegative as an expected shortfall under $\Qp$ of the random variable $J(  \cC+\cF-\ca)$, which is centered under $\Q$ and therefore under $\Qp$, by
\eqref{e:survdi}.
The first four formulas in \qr{L:discrs} directly follow from Definition \ref{d:ccr}  
and Lemma \ref{e:thelem}, which also implies that $\kva= \Ep\big(   h (\EC-\mathrm{KVA})^+ \big)= h (\EC-\mathrm{KVA})^+$. As $h$ is nonnegative, this KVA semilinear equation is equivalent to $$(\KVA>\EC  \mbox{ and } \KVA=0 ) \mbox{ or } (\KVA\le\EC  \mbox{ and } \KVA=\frac{h}{1+h} \EC ),$$
where $(\KVA>\EC  \mbox{ and } \KVA=0)$ contradicts the nonnegativity of $\EC$, whereas, for $h\in [0,1]$ as assumed and $\EC\ge 0$, $\KVA=\frac{h}{1+h} \EC$ implies $\kva\le \EC$, i.e.~$\max(\EC,\mathrm{KVA})=\EC$. This and Lemma \ref{e:thelem} yield
$$\fva=\Ep\Big( \gamma\big(\sum_b (\mtm_b-{\rm VM}_b) -\CA- \EC \big)^+ \Big)=\gamma\big(\sum_b (\mtm_b-{\rm VM}_b) -\CA- \EC\big)^+.$$
As
$\CA = 
\CCVA +\CMVA+\bcva+\bmva+\fva$, this is an FVA semilinear equation, which, as $\gamma$ is nonnegative, is equivalent to the FVA formula
$$\fva=\frac{\gamma}{1+\gamma}   \big(\sum_b (\mtm_b-{\rm VM}_b) -
%\ca
(\CCVA+\CMVA+
\bcva +\bmva)- \EC  \big)^+ .$$
Last, we have 
$\EC=\ES(J( \loss-\ca))$, where the identity $\loss-\ca=\cC-\CVA$ and the formula for $J(\cC-\CVA)$  in \qr{L:discrs} are obtained by
substituting the already derived XVA formulas in  
\eqref{e:recapprelccpc} and \eqref{e:recapprelccpf}.~\finproof 
 \brem 
The reason why funding disappears from the bank trading loss, i.e.~$J( \loss-\ca)=J( \cC-\cva)$, is because, in a one-period setup, 
the collateral borrowing requirements \eqref{eq:cF} of the bank are simply constants. Hence funding  triggers no risk to the bank, but only a deterministic cost. In a dynamic setup, funding generates both costs and risk.~\erem

\subsection{Extension to Several CCPs or CCP Services\label{ext:ccps}}

In the realistic case where the reference bank is a clearing member of several services of one or several CCPs, 
we index all the CCP related quantities in the above by an additional index $ccp$ in a  finite set disjoint from $I\cup C\cup B
%\cup E
$.
Then, with 
$\CA = 
\CCVA +\CMVA+\bcva+\bmva+\fva$ as before:
 
\bp\label{p:ccprm}
We have 
\bel
& \cC=  
\sum_{ccp, c}  (1- \So_c )(\cPo^{ccp}_c -\VMo^{ccp}_c -\IMo^{ccp}_c  )^+  +\sum_{ccp}\mu^{ccp} \cL ^{ccp}   \\&\qqq
+\sum_b  (1-\Ss_b)(\cP_b-{\rm VM}_b-\IM_b )^+, \mbox{where} \\ &\qqq\cL^{ccp} = \displaystyle\sum_{i}  (1-\S_i )
 \big( ( \mathcal{P}^{ccp}_i -\mtm^{ccp}_i -\IM^{ccp}_i  
 )^+ \\ &\qqq\qqq+ ( \cPs^{ccp}_i -\VMs^{ccp}_i -\PIMs_i^{ccp} )^+ -  \DF^{ccp}_i    \big)^+,
 \eel
 \bel
&\cF=  {\zeta}\sum_{ccp} \zeta\big(\IM^{ccp}+\PIMs^{ccp} + \DF^{ccp}\big)+ \\&
 \qqq
 + \zeta \sum_b  {  \IMs_b} +   \gamma   \big(\sum_b (\mtm_b-{\rm VM}_b) -\CA-\max(\EC,\kva) \big)^+,\\
 & \CA=\CCVA+\CMVA+\bcva +\bmva+\fva, \mbox{where}\\
& \qqq \CCVA
=  \Ep \Big[  \sum_{ccp, c}  (1- \So_c )(\cPo^{ccp}_c -\VMo^{ccp}_c -\IMo^{ccp}_c  )^+  +\sum_{ccp}
\mu^{ccp} \cL ^{ccp}  \Big]   , \\ 
&\qqq \CMVA 
= \sum_{ccp} \zeta\big(\IM^{ccp}+\PIMs^{ccp} + \DF^{ccp}\big) ,\\ 
  &
 \qqq  \bcva=\Ep\Big( \sum_b (1-\Ss_b)(\cP_b-{\rm VM}_b-\IM_b )^+   \Big) ,\\ 
   &\qqq \bmva=\zeta\sum_b  \IMs_b,\\
&\qqq \fva=\frac{\gamma}{1+\gamma} \Big( \big(\sum_b (\mtm_b-{\rm VM}_b) -
\\ &\qqq\qqq
(\CCVA+\CMVA+\bcva +\bmva
) -\EC  \big)^+   \Big), 
\\&\EC=\ES(J( \cC-\cva))\mbox{ and } \kva=\frac{h}{1+h} \EC, \mbox{where}
\\
 &\qqq \CVA=\CCVA+ \bcva  , 
\\
 & \qqq J( \cC-\cva )
 = J\Big( \sum_{ccp, c}(1-\So_c)(\cPo^{ccp}_c-\VMo^{ccp}_c-\IMo^{ccp}_c)^+    +\sum_{ccp}\mu^{ccp}\cL ^{ccp}  -  \CCVA \\&\qqq\qqq\qqq+
\sum_b (1-\Ss_b)(\cP_b-{\rm VM}_b-\IM_b )^+  -\bcva\Big).
\eel 
\ep
\proof In the case of several CCP services, the second line in \eqref{eq:cF} must be turned into 
$\sum_{ccp}\big( \IM^{ccp}+\PIMs^{ccp} +\DF^{ccp}\big)  +  \sum_b \IMs_b$;
the terms in the first lines of \eqref{e:recapprelccpc} and
\eqref{e:recapprelccpf}
must now be summed over the various CCP services in which the bank is involved as a clearing member. 
The rest of the analysis proceeds as before.
\finproof\\
 
Before passing to the case studies, we detail the calculation of economic capital under the member survival measure.
\bl \label{e:eclem} Denoting by 
$ q_{\alpha}$ 
the $\Qp$ value-at-risk, if $\Qp(  \cC-\cva =q_{\alpha})=0$, then
\beql{e:eccalc} 
&\EC=  \Es\big(  \cC-\cva \big|   \cC-\cva \geq q_{\alpha},J=1\big) .
\eeql 
\el
\proof If $\Qp(  \cC-\cva =q_{\alpha})=0$,
then, using equations (3.4) and (3.7) from \citeN{AT2002} to represent the tail mean expression\footnote{as per Definition 2.6 and further representation (3.7) from \citeN{AT2002}.} for the ES, we have by \eqref{L:discrs}:
\bel
\begin{array}{lcl}
\EC= \ES\big(J( \cC-\cva)\big) & = & \Ep\big(J( \cC-\cva)\big| J( \cC-\cva)\ge  q_{\alpha}\big))\\&=&\displaystyle\frac{\Ep\left(J( \cC-\cva)\mathds{1}_{\big\{J( \cC-\cva)\ge  q_{\alpha}\big\}}\right)}{\Qp\big(J( \cC-\cva)\ge  q_{\alpha}\big)}\\ 
&=&\displaystyle\frac{\Es\left(( \cC-\cva)\mathds{1}_{\big\{( \cC-\cva)\ge  q_{\alpha}\big\}}\mathds{1}_{\{J=1\}}\right)}{\Qs\big(( \cC-\cva)\ge  q_{\alpha},J=1\big)},
\end{array}
\eel 
using \eqref{e:survdi} on both numerator and denominator expressed in expectation form, which yields  \eqref{e:eccalc}.~\finproof

\section{Case Studies Setup\label{s:topract}}

We describe two possible applications of our XVA framework which will be illustrated by numerical case studies. To these ends, we introduce a market and credit model with parameters that can capture dependence between portfolio changes, joint defaults and possible averse exacerbated changes of the portfolio due to their owner default known as {wrong-way risk}. Then two networks will be defined to serve the numerical illustrations, one rather educational on the use of the XVA metrics and the other one reflecting the more realistic situation depicted by Figure \ref{fig:Network2CCPsUC}.

The CVA and KVA computations require a Monte-Carlo routine run under $\Qs$ in combination with a rejection technique in order to yield simulations under the survival measures associated with different clearing members, thereafter labeled CM* with * taking an identifier number.In the numerical applications that follow, all members play in turn the role of the reference bank in the theoretical XVA framework of Sections \ref{s:setup}-\ref{s:xvaa}.
For obtaining confidence intervals regarding  the expected shortfalls that are embedded in the KVA computations, the simulations are split into several batches, from which the mean of the estimated ES's yields the final ES estimate, while their standard deviation is used to define a confidence interval. 

The default time for member $i$ is generated based on Student-t copulas with correlated credit and market components, where credit components are reflected through the member's default times and the market components through their portfolio variation over the liquidation period following the default, proxied in our setup by the difference $\Delta \cP_i\eqdef  \cP_i -\mtm_i $. We denote by $\rho^{cr}>0$ the correlation coefficient of the copula Gaussian factor driving common defaults, by $\rho^{mkt}>0$ the correlation coefficient of the Gaussian factor driving common portfolio variations, and by ${\rho^{wwr}_i}>0$ the correlation coefficient of the Gaussian factor driving both portfolio variation and default for member $i$. The Student-t degree of freedom parameter is assumed to be the same for generating both members' default and portfolio variations. In equations, denoting by $F_{i}$ the marginal c.d.f. of member $i$'s default time and by $S_{\nu}$ the Student-t c.d.f. with degree of freedom $\nu$:
\beql{e:creditmarketcopula}
\left\{
\begin{array}{l}
      \tau_i=F_{i}^{-1}\Bigg({S_{\nu}}\left(\sqrt{\displaystyle\frac{\nu}{\mathcal{W}_i^{c}}}\left(\sqrt{\rho^{cr}} \mathcal{T}-\sqrt{{\rho^{wwr}_i}}\sqrt[4]{\displaystyle\frac{1-\rho^{cr}}{1-\rho^{mkt}}}\mathcal{X}_i+\right.\right.\\\left.\left. \qqq \sqrt{1-\rho^{cr}}\sqrt{1-\displaystyle\frac{{\rho^{wwr}_i}}{\sqrt{1-\rho^{cr}}\sqrt{1-\rho^{mkt}}}}\mathcal{T}_i\right)\right)\Bigg),\\
      \\
      \displaystyle\frac{\Delta \cP_i}{\Nom_i\sigma_i\sqrt{\Delta_{\ell}}}=   
    \sqrt{\displaystyle\frac{\nu}{\cW_i^{m}}}\left(\sqrt{\rho^{mkt}} \mathcal{E}+\sqrt{{\rho^{wwr}_i}}\sqrt[4]{\displaystyle\frac{1-\rho^{mkt}}{1-\rho^{cr}}}\mathcal{X}_i+\right. \\\left.  \qqq \qqq
      \sqrt{1-\rho^{mkt}}\sqrt{1-\displaystyle\frac{{\rho^{wwr}_i}}{\sqrt{1-\rho^{cr}}\sqrt{1-\rho^{mkt}}}}\mathcal{E}_i\right)
\end{array}
\right.
\eeql 
where $\Nom_i \in\mathbb{R}$ is a signed nominal of the portfolio of member $i$, $\sigma_i$ is its annualized relative volatility, $\Delta_{\ell}$ reflects a positive liquidation period accounting for the time taken by the CCP to novate or liquidate\footnote{cf. Section \ref{sec:OptNovDefMbb}.} defaulted portfolios, and
\begin{itemize}
    \item $\mathcal{T}$, $\mathcal{T}_i$, $\mathcal{E}$, $\mathcal{E}_i$ and $\mathcal{X}_i$ are i.i.d. standard normal random variables, where:
    \begin{itemize}
      \item $\mathcal{T}$ represents the common systemic factor for default times across members,
    \item $\mathcal{E}$ represents the common systemic factor for portfolio variations across members,
    \item $\mathcal{X}_i$ is the common factor co-driving portfolio variations and default time of member $i$,
    \item $\mathcal{T}_i$ is the idiosyncratic factor for member $i$'s default time,
    \item $\mathcal{E}_i$ is the idiosyncratic factor for member $i$'s portfolio variations;
    \end{itemize}
    \item $\cW_i^{c}$ and $\cW_i^{m}$ are i.i.d. random variables following $\chi^2$ distribution with degree of freedom $\nu$, independent from the above Gaussian random variables.
\end{itemize}
\brem 
In practice, margin computations rely on historical estimates based on several market stressed periods. Our approach, instead, aims at reflecting extreme market shocks with fat tailed  Student-t distributions of degree of freedom $\nu=3$, and volatility level within a reasonable range of $[20\%, 40\%]$.
Our static formulation depicts stationary increments of the defaulted portfolios' value over the liquidation period.
\erem
The above setup requires the following constraints on the correlation coefficients to be properly defined\footnote{otherwise, the model for both default time and portfolio variation factors is undefined due to their idiosyncratic coefficient term $\sqrt{1-\frac{{\rho^{wwr}_i}}{\sqrt{1-\rho^{cr}}\sqrt{1-\rho^{mkt}}}}$.}:
\beql{e:CorrelCond1}
\sqrt{1-\rho^{cr}}\sqrt{1-\rho^{mkt}}\geq {\rho^{wwr}_i}
\eeql

The "minus" sign in front of the common credit-market factor
$-\sqrt{{\rho^{wwr}_i}}$
for the default time component in \eqref{e:creditmarketcopula} ensures that the corresponding common factor accelerates defaults, whilst increasing the market exposure due to the $+\sqrt{{\rho^{wwr}_i}}$ factor in the second part of \eqref{e:creditmarketcopula}. 

In the examples that follow, market participants are identified by a number and can then be included in one of several of the considered CCPs.

\subsection{Single CCP Setup and initial XVA costs}\label{sec:SingleCCPUC}

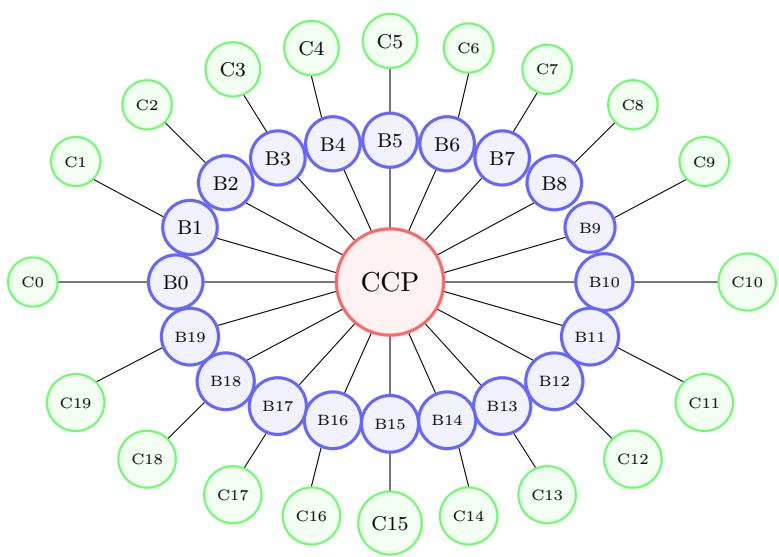
\begin{figure}[ht]
\begin{center}
\begin{tikzpicture}[
roundnodeC/.style={circle, draw=green!60, fill=green!5, thick, minimum size=5mm},
roundnodeB/.style={circle, draw=blue!60, fill=blue!5, very thick, minimum size=7mm},
roundnodeCH/.style={circle, draw=red!60, fill=red!5, very thick, minimum size=15mm},
scale=0.94, transform shape
]
%Nodes
\node at (0,0) [roundnodeCH]  (theCCP)    {CCP};
\node at (-3,0) [roundnodeB]   (bank1)     {\scriptsize B0};
\node at (-2.8,0.77) [roundnodeB]   (bank2)     {\scriptsize B1};
\node at (-2.3,1.4) [roundnodeB]   (bank3)     {\scriptsize B2};
\node at (-1.57,1.75) [roundnodeB]   (bank4)     {\scriptsize B3};
\node at (-0.8,1.95) [roundnodeB]   (bank5)     {\scriptsize B4};
\node at (0,2) [roundnodeB]   (bank6)     {\scriptsize B5};
\node at (0.8,1.95) [roundnodeB]   (bank7)     {\scriptsize B6};
\node at (1.57,1.75) [roundnodeB]   (bank8)     {\scriptsize B7};
\node at (2.3,1.4) [roundnodeB]   (bank9)     {\scriptsize B8};
\node at (2.8,0.77) [roundnodeB]   (bank10)     {\tiny B9};
\node at (3,0) [roundnodeB]   (bank11)     {\tiny B10};
\node at (2.8,-0.77) [roundnodeB]   (bank12)     {\tiny B11};
\node at (2.3,-1.4) [roundnodeB]   (bank13)     {\tiny B12};
\node at (1.57,-1.75) [roundnodeB]   (bank14)     {\tiny B13};
\node at (0.8,-1.95) [roundnodeB]   (bank15)     {\tiny B14};
\node at (0,-2) [roundnodeB]   (bank16)     {\tiny B15};
\node at (-0.8,-1.95) [roundnodeB]   (bank17)     {\tiny B16};
\node at (-1.57,-1.75) [roundnodeB]   (bank18)     {\tiny B17};
\node at (-2.3,-1.4) [roundnodeB]   (bank19)     {\tiny B18};
\node at (-2.8,-0.77) [roundnodeB]   (bank20)     {\tiny B19};

\node at (0,-3.4) [roundnodeC]  (client16)    {\scriptsize C15};

\node at (0,3.4) [roundnodeC]  (client6)    {\scriptsize C5};
\node at (-1.1,3.3) [roundnodeC]  (client5)    {\scriptsize C4};
\node at (-2.2,3) [roundnodeC]  (client4)    {\scriptsize C3};
\node at (-3.4,2.5) [roundnodeC]  (client3)    {\tiny C2};
\node at (-4.4,1.7) [roundnodeC]  (client2)    {\tiny C1};
\node at (-5,0) [roundnodeC]  (client1)    {\tiny C0};
\node at (1.1,3.3) [roundnodeC]  (client7)    {\tiny C6};
\node at (2.2,3) [roundnodeC]  (client8)    {\tiny C7};
\node at (3.4,2.5) [roundnodeC]  (client9)    {\tiny C8};
\node at (4.4,1.7) [roundnodeC]  (client10)    {\tiny C9};
\node at (5,0) [roundnodeC]  (client11)    {\tiny C10};

\node at (1.1,-3.3) [roundnodeC]  (client15)    {\tiny C14};
\node at (2.2,-3) [roundnodeC]  (client14)    {\tiny C13};
\node at (3.4,-2.5) [roundnodeC]  (client13)    {\tiny C12};
\node at (4.4,-1.7) [roundnodeC]  (client12)    {\tiny C11};

\node at (-1.1,-3.3) [roundnodeC]  (client17)    {\tiny C16};
\node at (-2.2,-3) [roundnodeC]  (client18)    {\tiny C17};
\node at (-3.4,-2.5) [roundnodeC]  (client19)    {\tiny C18};
\node at (-4.4,-1.7) [roundnodeC]  (client20)    {\tiny C19};

%Lines
\draw[-] (bank1.east) -- (theCCP.west);
\draw[-] (bank2.-22) -- (theCCP.170);
\draw[-] (bank3.-45) -- (theCCP.150);
\draw[-] (bank4.-45) -- (theCCP.130);
\draw[-] (bank5.-67) -- (theCCP.110);
\draw[-] (bank6.south) -- (theCCP.north);
\draw[-] (bank7.-113) -- (theCCP.70);
\draw[-] (bank8.-135) -- (theCCP.50);
\draw[-] (bank9.-135) -- (theCCP.30);
\draw[-] (bank10.-158) -- (theCCP.10);
\draw[-] (bank11.west) -- (theCCP.east);
\draw[-] (bank12.158) -- (theCCP.-10);
\draw[-] (bank13.135) -- (theCCP.-30);
\draw[-] (bank14.135) -- (theCCP.-50);
\draw[-] (bank15.113) -- (theCCP.-70);
\draw[-] (bank16.north) -- (theCCP.south);
\draw[-] (bank17.67) -- (theCCP.-110);
\draw[-] (bank18.45) -- (theCCP.-130);
\draw[-] (bank19.45) -- (theCCP.-150);
\draw[-] (bank20.22) -- (theCCP.-170);

\draw[-] (bank1.west) -- (client1.east);
\draw[-] (bank2.158) -- (client2.-45);
\draw[-] (bank3.135) -- (client3.-45);
\draw[-] (bank4.113) -- (client4.-67);
\draw[-] (bank5.113) -- (client5.south);
\draw[-] (bank6.north) -- (client6.south);
\draw[-] (bank10.22) -- (client10.-135);
\draw[-] (bank9.45) -- (client9.-135);
\draw[-] (bank8.67) -- (client8.-113);
\draw[-] (bank7.67) -- (client7.south);
\draw[-] (bank11.east) -- (client11.west);
\draw[-] (bank20.-158) -- (client20.45);
\draw[-] (bank19.-135) -- (client19.45);
\draw[-] (bank18.-113) -- (client18.67);
\draw[-] (bank17.-113) -- (client17.north);
\draw[-] (bank16.south) -- (client16.north);
\draw[-] (bank15.-67) -- (client15.north);
\draw[-] (bank14.-67) -- (client14.113);
\draw[-] (bank13.-45) -- (client13.135);
\draw[-] (bank12.-22) -- (client12.135);
\end{tikzpicture}
\end{center}
\caption{Financial network composed of 1 CCP, its 20 members (labeled by B) and one client per member}
\label{fig:Network1CCP20Mbs}
\end{figure}

We consider a single CCP service with 20 members labeled by $i\in 0\cdots \then=19$, only trading for cleared clients (i.e.~without bilateral or centrally cleared proprietary trading).  Each member faces one client. The ensuing financial network is depicted by \hyperref[fig:Network1CCP20Mbs]{Figure \ref{fig:Network1CCP20Mbs}}.

All clients are assumed to be risk-free.
For any member $i$, its posted IM to the CCP is calculated based on the idea of a VM call not fulfilled over a time period $\Delta_s<\Delta_{\ell}$ at a confidence level $\alpha\in(1/2,1)$, using a VaR metric\footnote{under the member survival measure.} applied to the non-coverage of VM call taken also to follow a scaled Student-t distribution $\mathcal{S}_{\nu}$ with $\nu$ degrees of freedom:  
    \beql{e:IMCaldDef}
    \IM_i=\VaR_{1-\alpha}\big(\Nom_i\sigma_i\sqrt{\Delta_s}\mathcal{S}_{\nu}\big)=|\Nom_i|\,\sigma_i\sqrt{\Delta_s}{S_{\nu}}^{-1}(\alpha)
    \eeql
    where ${S_{\nu}}^{-1}$ is the inverse c.d.f. of a Student-t distribution with degree of freedom $\nu$. 
The default fund is calculated at the CCP level as
\beql{e:DF_size_Cover2}
{\rm Cover2} = \SLOIM_{(0)} + \SLOIM_{(1)},
\eeql
for the two largest stressed losses over IM ($\SLOIM_i$) among members, identified with subscripts $(0)$ and $(1)$, where  SLOIM is calculated as the value-at-risk at confidence level $\alpha'>\alpha$ of the loss over IM, i.e.
\beql{e:SLOIMCaldDef}
\SLOIM_i=\VaR_{\alpha'}\big(\Nom_i\sigma_i\sqrt{\Delta_s}\mathcal{S}_{\nu}-\IM_i\big)= |\Nom_i|\,\sigma_i\sqrt{\Delta_s}\Big({S_{\nu}}^{-1}(\alpha')-{S_{\nu}}^{-1}(\alpha)\Big).
\eeql
The total amount \eqref{e:DF_size_Cover2} is then allocated between the clearing members to define their (funded) default fund contributions as $\DF_i=\displaystyle\frac{\SLOIM_i}{\sum_{j}\SLOIM_j}{\rm Cover2}$. 

The  $\Nom_j$'s of other clearing members are not observable by a given one. However, following \citeN{MNW14} and \citeN{Li18}, $\Nom_{(i)}$ denoting the $i$-th largest absolute nominal amount for $i\in 0\cdots \then=19$, a parameterization of the form
\beql{e:expPflioSize}
\Nom_{(i)} = \alpha e^{-\beta (i+1)},\qqq \alpha,\,\beta>0
\eeql
can be fit to the total default fund held by the CCP\footnote{item referenced as 4.3.15 in \citeN{CPMIIOSCO15}, {\it Value of pre-funded default resources (excluding initial and retained variation margin) held for each clearing service in total, post-haircut.} in the quantitative disclosure documents.} and the sum of its five largest default fund contributions\footnote{item referenced as 18.4.2 in \citeN{CPMIIOSCO15}:{\it For each segregated default fund with 25 or more members; Percentage of participant contributions to the default fund contributed by largest five clearing members in aggregate.}; or item referenced 18.4.1 for CCP services with less than 25 members}, published each quarter for most of the CCPs and that are public data. The inferred parameter $\alpha$ and $\beta$ from the default fund data are used to depict a similar pattern on the nominal sizes\footnote{as if the default fund amounts are proportional to the portfolio sizes.}. 
The participants and portfolios parameter inputs are detailed in \hyperref[tab:MbNework1CCP20MbSetup]{Table \ref{tab:MbNework1CCP20MbSetup}}, where id is the identifier of the CM, DP stands for the one year probability of default of the member expressed in percentage points, size represent the overall portfolio size of the member detained within the CCP, and vol is the annual volatility used for the portfolio variations. 

\begin{center}
\begin{adjustbox}{width=\textwidth,center=\textwidth,captionbelow={Member characteristics and portfolio parameters, ordered by decreasing member size.},
label={tab:MbNework1CCP20MbSetup},float=table}
\begin{tabular}{c}
    \begin{tabular}{ |l|x{0.9cm}|x{0.9cm}|x{0.9cm}|x{0.9cm}|x{0.9cm}|x{0.9cm}|x{0.9cm}|x{0.9cm}|x{0.9cm}|x{0.9cm}| } 
    \hline cm id & 0 & 1 & 2 & 3 & 4 & 5 & 6 & 7 & 8 & 9 \\ \hline
    DP (\%) & 0.5 & 0.6 & 0.7 & 0.8 & 0.9 & 2 & 1.9 & 1.8 & 1.7 & 1.6\\ \hline
    size & -242 & 184 & 139 & 105 & -80 & -61 & -46 & 35 & 26 & -20 \\ \hline
    vol (\%) & 20 & 21 & 22 & 23 & 24 & 25 & 26 & 27 & 28 & 29 \\ \hline
    \end{tabular}
    \\ \\
    \begin{tabular}{ |l|x{0.9cm}|x{0.9cm}|x{0.9cm}|x{0.9cm}|x{0.9cm}|x{0.9cm}|x{0.9cm}|x{0.9cm}|x{0.9cm}|x{0.9cm}| } 
    \hline cm id & 10 & 11 & 12 & 13 & 14 & 15 & 16 & 17 & 18 & 19\\ \hline
    DP (\%) & 1.5 & 1.4 & 1.3 & 1.2 & 1.1 & 1 & 0.9 & 0.8 & 0.7 & 0.6\\ \hline
    size & -15 & -11 & -9 & -6 & 5 & -4 & -3 & 2 & 2 & -1\\ \hline
    vol (\%) & 30 & 31 & 32 & 33 & 34 & 35 & 36 & 37 & 38 & 39\\ \hline
    \end{tabular}
\end{tabular}
\end{adjustbox}
\end{center}

The portfolios listed in the Table \ref{tab:MbNework1CCP20MbSetup} relates to the members towards the CCP (which are mirroring the ones between the members and their clients). The sizes sum up to 0, in line with the CCP clearing condition (first identity in Assumption \ref{h:contract}, here without proprietary trades).

The parameters of the XVA costs calculations are summarized in \hyperref[tab:XVAconfig1CCP20Mbs]{Table \ref{tab:XVAconfig1CCP20Mbs}}. Note that the chosen period length of $T=5$ years covers the bulk (if not the final maturity) of most realistic CCP portfolios.
\begin{table}[ht]
\begin{center}
\begin{tabular}{ |l|c| }
 \hline
 One-period length $T$ & 5 years \\ 
 \hline
 Liquidation period at default $\Delta_l$ & 5 days \\
 \hline
 Portfolio variations correlation $\rho^{cr}$'s & 30\%\\
 \hline
 Credit factors correlation $\rho^{mkt}$'s  & 20\% \\ \hline
 Correlation between credit factors and portfolio variations ${\rho^{wwr}_i}$'s&  20\% \\ \hline 
 IM covering period (MPoR) & 2 days\\ \hline
 IM quantile level & 95\% \\ \hline
Funding blending ratio $\zeta/\gamma$ 
 & 25\% \\
 \hline
 SLOIM calculation\footnote{such confidence level at $97\%$ for SLOIM in DF calibration allows for a ratio of default fund over initial margin of about $10\%$ in our calculations, a ratio (of this level or less) often observed in practice.}
 for DF Cover-2 & VaR 97\%\\
 \hline
 DF allocation rule & based on IM\\
 \hline
 Quantile level used for clearing members EC calculation & 99.75\%\\
 \hline
 Hurdle rate $h$ used for KVA computations & 10.0\%\\
 \hline
 Number of Monte-Carlo simulation (for CCVA \& KVA computations) & 5,000,000 \\ \hline
 Number of batches (for KVA computations) & 50 \\ \hline
\end{tabular}
\caption{XVAs calculation configuration}
\label{tab:XVAconfig1CCP20Mbs}
\end{center}
\end{table}

For each member, the CCVA, CMVA and KVA costs are calculated and reported in \hyperref[tab:InitXVAsCosts]{Table \ref{tab:InitXVAsCosts}}. For KVA, two calculations have been performed, one based on ES at $99^{th}$ percentile level and another one based on $99.75^{th}$ percentile level. The amount in square bracket is the corresponding quantile level from which average is calculated and numbers in parenthesis represent the 95\% confidence interval in relative difference from calculated metric for both CCVA and KVA. All the XVA numbers decrease with the member size.
\begin{center}
\begin{adjustbox}{width=0.8
\textwidth,center=\textwidth,captionbelow={Initial XVA costs:  estimates, [value-at-risk underlying the KVA estimate] and (95\% confidence level errors)},label={tab:InitXVAsCosts},float=table}
\begin{tabular}{ |c|c|c|c|c| }
\hline
cm id & CMVA & CCVA & KVA (99\%) & KVA (99.75\%)\\ \hline
0 & 0.0687 & 0.0442 (0.6\%) & 0.2093 [0.0948] (1\%) & 0.4142 [0.2306] (1.6\%)\\ \hline
1 & 0.0656 & 0.0412 (0.7\%) & 0.2087 [0.0855] (1.2\%) & 0.437 [0.2267] (1.8\%)\\ \hline
2 & 0.0604 & 0.0327 (0.7\%) & 0.1683 [0.0679] (1.2\%) & 0.355 [0.1823] (1.6\%)\\ \hline
3 & 0.0544 & 0.026 (0.7\%) & 0.1355 [0.0543] (1.2\%) & 0.2863 [0.1466] (1.6\%)\\ \hline
4 & 0.0485 & 0.0191 (0.8\%) & 0.103 [0.0395] (1.2\%) & 0.2224 [0.1112] (1.7\%)\\ \hline
5 & 0.0834 & 0.0133 (0.8\%) & 0.0809 [0.0295] (1.2\%) & 0.1772 [0.0886] (1.6\%)\\ \hline
6 & 0.0623 & 0.0111 (0.8\%) & 0.0667 [0.0251] (1.1\%) & 0.1439 [0.0728] (1.6\%)\\ \hline
7 & 0.0467 & 0.0099 (0.7\%) & 0.0557 [0.0223] (1.1\%) & 0.1171 [0.0606] (1.5\%)\\ \hline
8 & 0.0341 & 0.0078 (0.7\%) & 0.0432 [0.0174] (1.1\%) & 0.091 [0.0471] (1.6\%)\\ \hline
9 & 0.0256 & 0.006 (0.8\%) & 0.0342 [0.0137] (1.2\%) & 0.0721 [0.0371] (1.7\%)\\ \hline
10 & 0.0187 & 0.0048 (0.8\%) & 0.0266 [0.0107] (1.1\%) & 0.0561 [0.0289] (1.6\%)\\ \hline
11 & 0.0132 & 0.0037 (0.7\%) & 0.0202 [0.0081] (1.1\%) & 0.0423 [0.0219] (1.5\%)\\ \hline
12 & 0.0104 & 0.0031 (0.7\%) & 0.017 [0.0069] (1.2\%) & 0.0358 [0.0185] (1.6\%)\\ \hline
13 & 0.0066 & 0.0022 (0.7\%) & 0.0116 [0.0047] (1.1\%) & 0.0244 [0.0127] (1.6\%)\\ \hline
14 & 0.0052 & 0.0019 (0.7\%) & 0.01 [0.0041] (1.1\%) & 0.021 [0.0108] (1.6\%)\\ \hline
15 & 0.0039 & 0.0016 (0.7\%) & 0.0082 [0.0033] (1.1\%) & 0.0172 [0.0089] (1.5\%)\\ \hline
16 & 0.0027 & 0.0012 (0.7\%) & 0.0063 [0.0026] (1.1\%) & 0.0132 [0.0068] (1.6\%)\\ \hline
17 & 0.0017 & 0.0008 (0.7\%) & 0.0043 [0.0017] (1.1\%) & 0.009 [0.0047] (1.5\%)\\ \hline
18 & 0.0015 & 0.0009 (0.7\%) & 0.0044 [0.0018] (1.1\%) & 0.0092 [0.0048] (1.6\%)\\ \hline
19 & 0.0007 & 0.0004 (0.7\%) & 0.0023 [0.0009] (1.1\%) & 0.0047 [0.0025] (1.5\%)\\ \hline
\end{tabular}
\end{adjustbox}
\end{center}

\begin{figure}[ht]
\begin{center}
\includegraphics[width=1\textwidth]{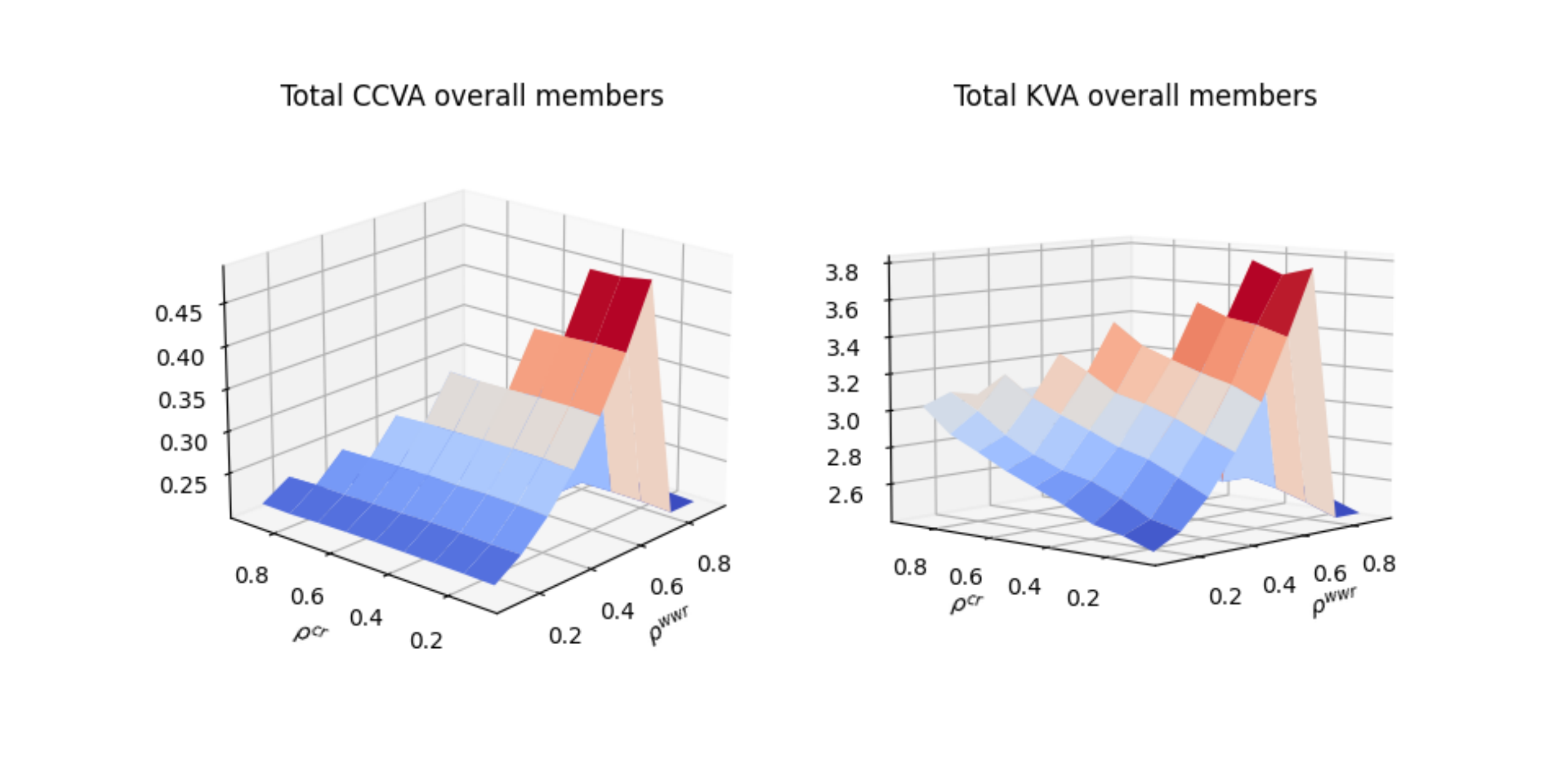}
\end{center}
\caption{CCVA and KVA w.r.t. credit factors correlation and credit and portfolio variation factors correlation}
\label{fig:CCVA_KVA_wrt_correl}
\end{figure}

\begin{figure}[ht]
\begin{center}
\includegraphics[width=1\textwidth]{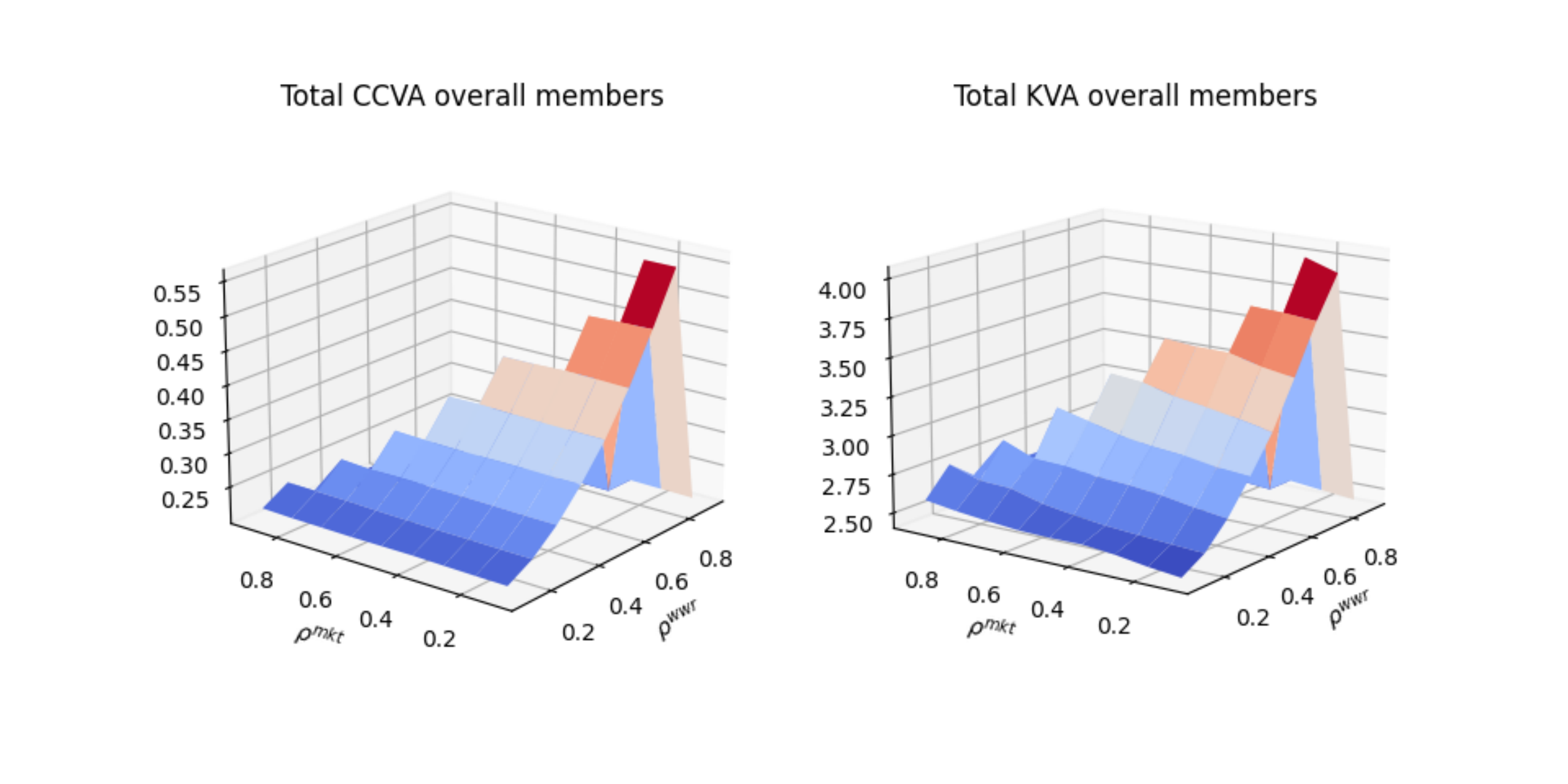}
\end{center}
\caption{CCVA and KVA w.r.t. market factors correlation and credit and portfolio variation factors correlation}
\label{fig:CCVA_KVA_wrt_mkt_mkt_correl}
\end{figure}

To assess the average behavior w.r.t. 
$\rho^{cr}$, $\rho^{mkt}$ and ${\rho^{wwr}}$ of the CCVA and KVA, we vary these correlations between 10\% and 90\% and display in Figures \ref{fig:CCVA_KVA_wrt_correl} and \ref{fig:CCVA_KVA_wrt_mkt_mkt_correl} the corresponding metrics, aggregated over all clearing members successively considered as the reference bank. As expected, the KVA depicts both an increase with respect to $\rho^{cr}$, $\rho^{mkt}$ and $\rho^{wwr}$, though ${\rho^{wwr} }$ has more impact than $\rho^{cr}$ and $\rho^{mkt}$ (right panels in Figures \hyperref[fig:CCVA_KVA_wrt_correl]{\ref{fig:CCVA_KVA_wrt_correl}} and \hyperref[fig:CCVA_KVA_wrt_mkt_mkt_correl]{\ref{fig:CCVA_KVA_wrt_mkt_mkt_correl}}). As seen on the left panels of Figures \hyperref[fig:CCVA_KVA_wrt_correl]{ \ref{fig:CCVA_KVA_wrt_correl}}  and \hyperref[fig:CCVA_KVA_wrt_mkt_mkt_correl]{\ref{fig:CCVA_KVA_wrt_mkt_mkt_correl}}, there are very marginal changes for CCVA w.r.t. $\rho^{cr}$ and $\rho^{mkt}$, but a significant impact of $\rho^{wwr}$. This is understandable as, apart for modulations of the measure with respect to which each individual CCVA is assessed, the CCVA aggregated over clearing members is essentially an expectation of the CCP loss $\cL$ (cf. the second line of \eqref{L:discrs}). The individual CCVAs  (as per the first line of \eqref{L:discrs}) of each clearing member, however, may depend on 
$\rho^{cr}$ and $\rho^{mkt}$ (on top of of $\rho^{wwr}$) in a strong and nontrivial manner, via the allocation coefficient $\mu$.

\subsection{Two CCPs Network Setup}
We consider the case of \hyperref[fig:Network2CCPsUC]{Figure \ref{fig:Network2CCPsUC}} where there are two CCPs with some common members and stress test is considered from the perspective of one of these common members. The motivation for this case is to provide a realistic example mimicking in a simplified way the trading firm Ronin Capital, which had memberships on both FICC GSD\footnote{Government Securities Division} segments, hereafter denominated by CCP1 and CME Futures, hereafter denominated by CCP2 in March 2020. It is well known that a VaR type risk measure is not sub-additive, in particular for credit portfolios as illustrated in Example 5.4 in \citeN{AT2002} and Example 2.25 in \citeN{McNeilFreyEmbrechts2015} for a portfolio of defaultable bonds, so that for a common member adding VaR estimates of trading losses on two CCPs separately can lead to underestimated levels with respect to the actual VaR of the global exposition of the member. As such, stress test exercises accounting for common memberships could reveal a larger risk compared to the exercise where stress tests are conducted separately on each CCP. 

\begin{figure}[ht]
    \begin{minipage}{0.48\textwidth}
    \centering
    \includegraphics[width=1\textwidth]{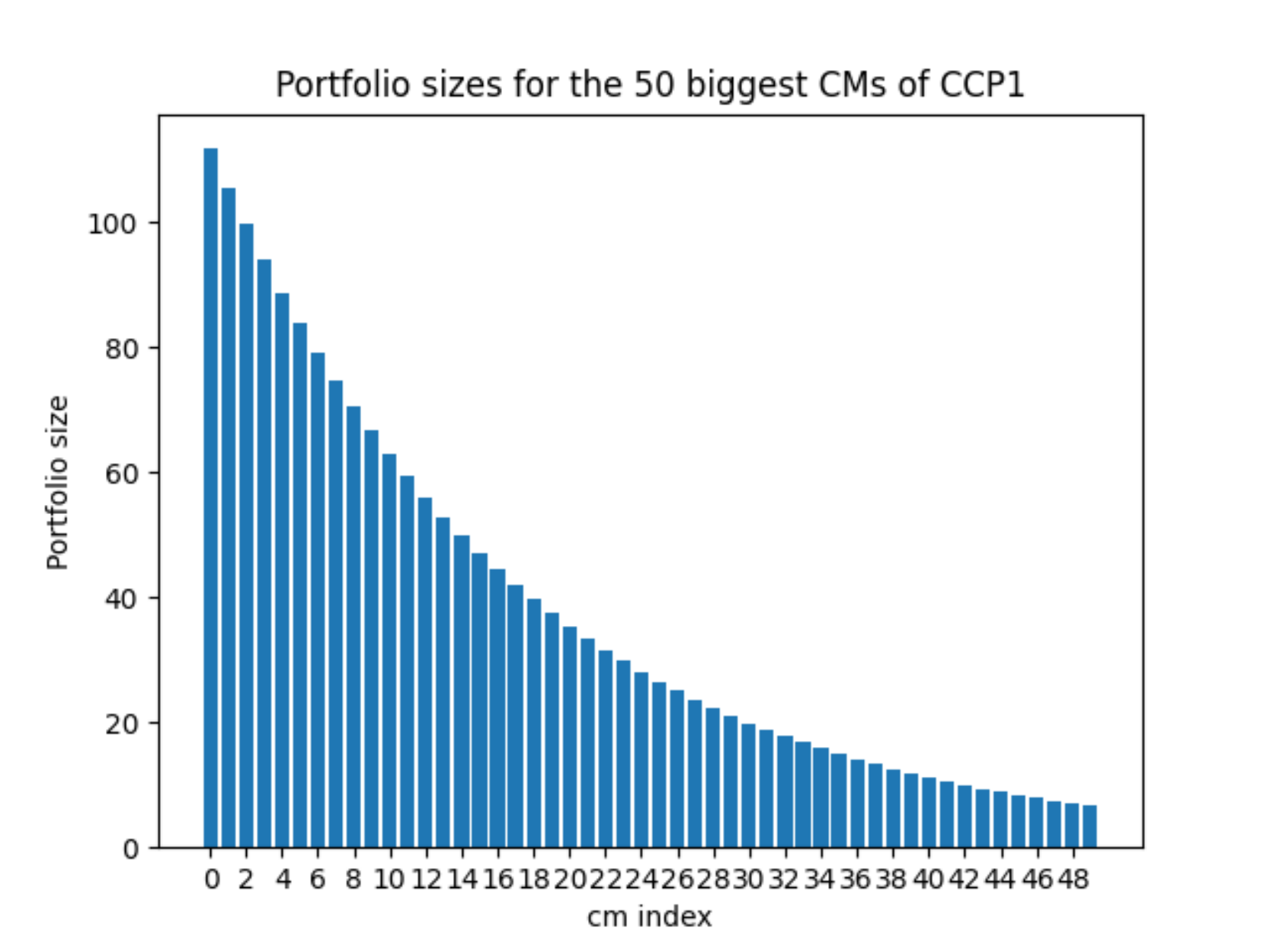}
    \caption{Decreasing absolute nom$_i$ per member for   CCP1}\label{Fig:DFPropPerCMCCP1_155Mbs}
    \end{minipage}\hfill
    \begin{minipage}{0.48\textwidth}
    \centering
    \includegraphics[width=1\textwidth]{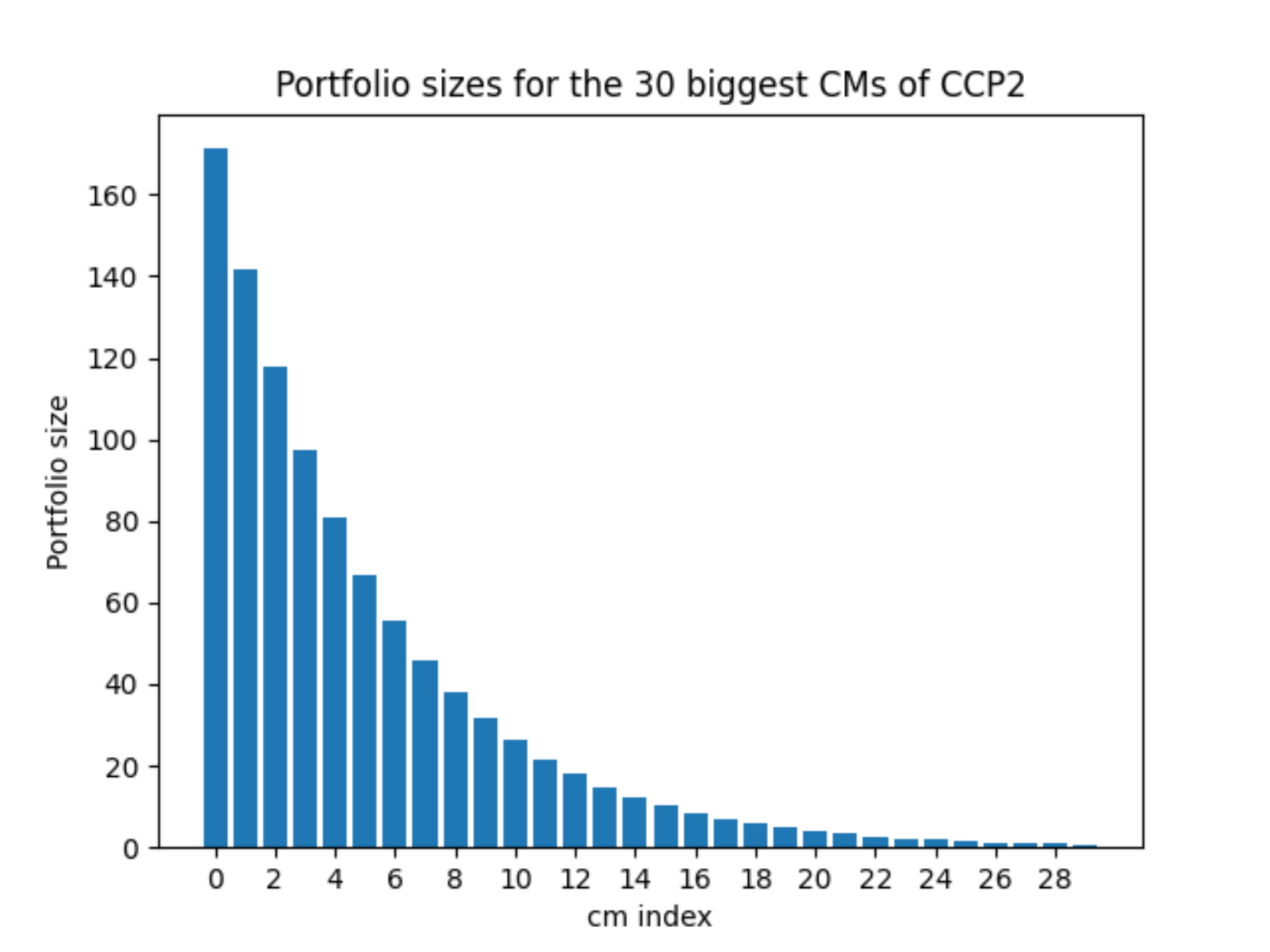}
    \caption{Decreasing absolute nom$_i$ per member for CCP2}    \label{Fig:DFPropPerCMCCP2_56Mbs}
    \end{minipage}
\end{figure}

To perform the analysis, the following setup is considered:
\begin{itemize}
    \item all members have only clearing client positions\footnote{Ronin Capital had in fact only a house account and was thus not clearing any client position.}, with 123 members on CCP1 and 56 members on CCP2, out of which 24 are common to both CCPs,
    \item all clients are assumed default free,
    \item both CCPs use configuration as per Table \ref{tab:XVAconfig1CCP20Mbs},
    \item the sizes of the positions are assumed exponentially distributed in the sense that from the most exposed member to the least one, absolute value of positions decrease exponentially with the form in \eqref{e:expPflioSize} as depicted by Figures \ref{Fig:DFPropPerCMCCP1_155Mbs} and \ref{Fig:DFPropPerCMCCP2_56Mbs} respectively,
    \item the proportion of the default fund detained by the 5 biggest members is $25\%$ for CCP1 and $61\%$ for CCP2\footnote{taken from the quantitative disclosure of both CCPs as of third quarter of 2020.},
    \item the size of the default fund of CCP1 is assumed to be twice the one of the default fund of CCP2.
\end{itemize}
All data used are either public or have been anonymized. Similar configuration as given in \hyperref[tab:XVAconfig1CCP20Mbs]{Table \ref{tab:XVAconfig1CCP20Mbs}} is used, apart from the number of Monte-Carlo simulations reduced to 2 millions for memory capacity reasons.

The clearing conditions are ensured by setting the sum of the portfolio sizes nom$_i$ to zero  on each CCP. The situation of member 3, exposed to both CCPs, as the defaulting member, corresponds roughly to the situation of Ronin Capital in 2018. In particular, an annual  probability of default of 0.1\% corresponds roughly to a BBB rating, that was assigned to Ronin Capital in 2018 for its issuances\footnote{\url{https://www.spglobal.com/marketintelligence/en/news-insights/blog/banking-essentials-newsletter-july-edition-2}}.

\section{Stress Test Exercises}\label{ss:stexo}
As outlined in the capital requirements regulation detailed in \citeN{EU2013} article 290, financial institutions must conduct regular stress test exercises of their credit and counterparty exposures. Paragraph 8 of this article also stipulates the reverse stress test\footnote{see dedicated definition p.12 in \citeN{FED2012} and articles 97, 98 p. 37 in \citeN{ECB2018} for official regulatory definitions.} requirement to 
\begin{quote}
    [...] identify extreme, but plausible, scenarios that could result in significant adverse outcomes.
\end{quote}
{
This is complemented by article 302 on the exposure financial institutions may have towards CCPs:
\begin{quote}
    Institutions shall assess, through appropriate scenario analysis and stress testing, whether the level of own funds held against exposures to a CCP, including potential future credit exposures, exposures from default fund contributions and, where the institution is acting as a clearing member, exposures resulting from contractual arrangements as laid down in Article 304, adequately relates to the inherent risks of those exposures.    
\end{quote}
}

In practice, stress test exercises aim at assessing the capacity of financial institutions to absorb financial and economic shocks. In regular exercises, such as the ones conducted by the European Banking Authority, the shocks are usually considered under so called {\it central} and {\it baseline} macro-economic scenarios corresponding to a median quantile and {\it adverse} scenario usually taken as a $90^{th}$ percentile reflecting severe yet plausible scenario that can occur once every 10 years\footnote{such confidence levels are suggested by the Federal Reserve outlining p.10 in \citeN{FED2013} the various recession periods of the United States listed in their Table 1 p. 14. The 2021 instructions in \citeN{EBA21} also indicate p.72 that stressed market risk factors are based on shocks specified in \citeN{ESRB2020}, citing
\citeN[p. 29]{FED17}, with the US recessions periods as stressful economic episodes.}. Additionally, extreme scenarios can be considered for measuring the capital adequacy\footnote{cf.~paragraph 5.1 p.11 from \citeN{BCBS05}.} 
for absorbing extremely severe losses around confidence level at $99.9\%$. From a clearing member perspective, this requires to have the capacity of scanning certain points of its trading loss distribution. In our framework, this boils down to identifying particular levels of the distribution of the trading loss $\cC=\CVA=\cC-\CCVA-\bcva$ of the reference clearing member bank, where the different terms are detailed in Proposition \ref{p:ccprm}.\smallskip

The other type of stress test exercises, referenced as {\it reverse stress test}\footnote{see also dedicated definition on p.12 in \citeN{FED2012} and articles 97, 98 p. 37 in \citeN{ECB2018} for official regulatory definitions.} \citep{BelliniEichhornMayenberger20book}, consists in identifying the probability of reaching a given loss level as well as describing the scenario configuration such as projected defaults and loss magnitude leading to such loss levels. The distribution must span a sufficient large spectrum of losses, including the ones targeted by the exercise, but it also has to be sufficiently rich numerically to allow identifying combinations of events leading to such losses. 

Confidence intervals of corresponding extreme scenario probabilities should complement the analysis to ensure the reliability of the used model and numerical methods.

Regulators have the ability to challenge financial institutions on these elements and demand for improvements\footnote{this may entail re-assessment of the Pillar 2 guidance additional capital requirement set in the annual Supervisory Review and Evaluation Process reported by Banks, cf. \citeN{ECBPG21} for a brief definition and use and \citeN{BIS-P2G-19} for more extensive details as well as \citeN{FRB20} for similar requirements.}.

We now briefly explain how to identify and exploit the scenarios leading to contribute the most to economic capital, in the spirit of \citeN{AlbaneseCrepeyIabichino2020a}. 
We denote by $M$ the number of Monte-Carlo scenario for which $J=1$, i.e.~survival of the reference bank. Its trading loss $\cC-\cva$ for a simulation $m$ is given by $\cC^{m}-\cva$, where  $m\in 1\cdots M$ enumerates the simulated scenarios for which the reference member bank ends up in survival state. 

To get an estimate of the economic capital based on expected shortfall, relying on \citeN[Definition 2.6 and Proposition 4.1]{AT2002}, we calculate, for a high confidence level $\alpha\in (\frac{1}{2},1)$ and $[x]$ denoting the integer part of any $x\in\mathbb{R}$,
\beql{e:EstimES}
\widehat{\ES}\left(\cC-\cva\right):=\displaystyle\frac{1}{M- [\alpha M] }\sum_{m=[\alpha M]+1}^{M}\left\{\cC^{(m)}-\cva\right\},
\eeql
where the $\cC^{(m)}-\cva$'s are the simulated trading losses of the reference bank ranked in increasing order.

To obtain the contribution of any simulated scenario $m'$ (with $(m')\ge [\alpha M]$) to the economic capital estimated by \eqref{e:EstimES}, 
i.e.
\beql{e:EstimESwoSc}
\begin{array}{l}
     \widehat{\ES}^{- m'}\left(\cC-\cva\right):= \\ \\
     \hspace{1cm}\displaystyle\frac{1}{M-1-[\alpha (M-1)]}\left\{\big(M-[\alpha M]\big)\widehat{\ES}\left(\cC-\cva\right)-\left(\cC^{ m' }-\cva\right)\right\}.
\end{array}
\eeql
Yhe contribution $\delta_{ m' }\widehat{\ES}\left(\cC-\cva\right)$ of scenario $ m' $ to $\widehat{\ES}\left(\cC-\cva\right)$ is then given by:
\beql{e:ESscContrib}
\delta_{ m' }\widehat{\ES}\left(\cC-\cva\right)
=
\widehat{\ES}\left(\cC-\cva\right)-\widehat{\ES}^{- m'}\left(\cC-\cva\right)
\eeql
 
To illustrate the various flavors of stress test exercises that can be conducted by a CCP member, we report numerical results for the two network examples introduced in Section \ref{s:topract}. We start with a reverse stress test exercise on example covered by \hyperref[tab:MbNework1CCP20MbSetup]{Table \ref{tab:MbNework1CCP20MbSetup}}. For this first illustration, a specific extreme loss is targeted and  the corresponding probability of loss reaching at least such target level is estimated. We then consider the example illustrated by \hyperref[fig:Network2CCPsUC]{Figure \ref{fig:Network2CCPsUC}} where projected loss levels for specific confidence levels are indicated for the members with common memberships on the two CCPs. 

\subsection{Numerical Results\label{ss:numres}}

In \hyperref[tab:STexample]{Table \ref{tab:STexample}}, we report, for the example summarized in \hyperref[tab:MbNework1CCP20MbSetup]{Table \ref{tab:MbNework1CCP20MbSetup}}, the $99.9^{th}$ percentile trading loss levels, referenced as extreme quantile, with corresponding (asymmetric) confidence intervals based on the approach proposed in \citeN[Section G.2]{MHE2017}.
This is done for every clearing member
successively playing the role of the reference bank in the setup of Sections \ref{s:setup}-\ref{s:xvaa}. We also compute the probabilities of reaching a loss equal to 1.5 times the obtained extreme quantile level, referenced as RST scenario, with corresponding confidence levels\footnote{the calculation of the latter confidence intervals of the probability of being above a quantile relies on the same numerical approach based on batches used for KVA calculations. Also, the batch approach leads to reasonably tight confidence intervals for the RST scenario probabilities.}.
\begin{center}
\begin{adjustbox}{width=0.7\textwidth,center=\textwidth,captionbelow={Stress test (ST) extreme quantile, $1.5\times$ ST extreme quantile and RST probability to breach 1.5 times the $99.9^{th}$ quantile loss level, for each member, based on $5,000,000$ simulations  (in parentheses: corresponding 95\% confidence intervals).},label={tab:STexample},float=table}
\begin{tabular}{ |c|c|c|c| }
\hline
cm id & $99.9\%$  & $1.5\times\,\,99.9\%$   & RST scenario probability\\ \hline
0 & 3.9949 (-1.1\%, 1.4\%) & 5.9924 & 0.0387\% (5\%)\\ \hline
1 & 4.1141 (-1.3\%, 1.7\%) & 6.1712 & 0.0428\% (4.4\%)\\ \hline
2 & 3.3409 (-1.6\%, 1.4\%) & 5.0114 & 0.0435\% (4.4\%)\\ \hline
3 & 2.6914 (-1.4\%, 1.6\%) & 4.0371 & 0.0437\% (4.3\%)\\ \hline
4 & 2.0695 (-1.2\%, 1.8\%) & 3.1043 & 0.0452\% (4.5\%)\\ \hline
5 & 1.6715 (-1.7\%, 1.5\%) & 2.5073 & 0.044\% (4.6\%)\\ \hline
6 & 1.3517 (-1.3\%, 1.6\%) & 2.0276 & 0.0437\% (3.9\%)\\ \hline
7 & 1.1032 (-1.4\%, 1.7\%) & 1.6548 & 0.0434\% (4.2\%)\\ \hline
8 & 0.8554 (-1.5\%, 1.5\%) & 1.2831 & 0.0435\% (4.1\%)\\ \hline
9 & 0.6817 (-1.7\%, 1.5\%) & 1.0226 & 0.0433\% (4.2\%)\\ \hline
10 & 0.5287 (-1.7\%, 1.4\%) & 0.7931 & 0.0429\% (4.4\%)\\ \hline
11 & 0.4001 (-1.7\%, 1.4\%) & 0.6002 & 0.0427\% (4.4\%)\\ \hline
12 & 0.3385 (-1.3\%, 1.5\%) & 0.5078 & 0.0431\% (4.2\%)\\ \hline
13 & 0.2314 (-1.5\%, 1.5\%) & 0.3471 & 0.0428\% (4.6\%)\\ \hline
14 & 0.1972 (-1.3\%, 1.6\%) & 0.2958 & 0.0429\% (4.4\%)\\ \hline
15 & 0.1623 (-1.4\%, 1.5\%) & 0.2435 & 0.0427\% (4.3\%)\\ \hline
16 & 0.1249 (-1.4\%, 1.5\%) & 0.1874 & 0.0427\% (4.4\%)\\ \hline
17 & 0.0852 (-1.5\%, 1.4\%) & 0.1278 & 0.0434\% (4\%)\\ \hline
18 & 0.0873 (-1.4\%, 1.6\%) & 0.131 & 0.0427\% (4.1\%)\\ \hline
19 & 0.0447 (-1.6\%, 1.3\%) & 0.0671 & 0.0431\% (4\%)\\ \hline
\end{tabular}
\end{adjustbox}
\end{center}
Our description of the scenarios leading to such losses includes the identified defaulted members, the generated losses and the allocated loss coefficient of the reference clearing member (CM1 in this example).  \hyperref[tab:RSTscDetailsexample]{Table \ref{tab:RSTscDetailsexample}} provides the description of the 20 worst scenarios, contributing the most to the EC estimation for the second biggest member, that is CM1\footnote{its theoretical number of scenarios above the RST loss level should be 2076, i.e. the number of MC simulations of 5,000,000 multiplied by CM1's survival probability over 5 years and by CM1's RST loss level probability estimated in \hyperref[tab:STexample]{Table \ref{tab:STexample}} as 0.0428\%, 
which is of course far too many to report. Nonetheless a focus on the 20 worst ones already illustrates the type of information that can be exploited for such exercises.}. Most of these scenarios are driven by significant losses stemming from CM0's default, reflecting the highly concentrated position of CM0. Note that the $16^{th}$ worst loss scenario for member 1 entails 8 defaults, including the one of CM0, which is the only one to generate losses beyond its posted margin (i.e. to trigger a loss to the surviving members).

\begin{center}
\begin{adjustbox}{width=0.9\textwidth,center=\textwidth,captionbelow={Economic Capital 20 worst scenarios details for member 1 in decreasing order of total loss where column with header $\mu$ indicates allocated coefficient loss to member 1 and $n$ is the number of defaults within the scenario.},label={tab:RSTscDetailsexample},float=table}
\begin{tabular}{ |c|c|c|c|c|c| }
\hline
Rank & Total Loss & n & 
$\hspace{0.4cm}\mu\hspace{0.4cm}$ & Defaulters & Losses triggered by defaulters \\ \hline
1 & 10.97 & 3 & 0.22 & cm0, 9, 16 & 544.37, 0, 0 \\ \hline
2 & 8.18 & 3 & 0.23 & cm0, 7, 11 & 391.38, 0, 0.55 \\ \hline
3 & 7.84 & 3 & 0.24 & cm0, 6, 7 & 356.66, 0, 0 \\ \hline
4 & 6.74 & 1 & 0.21 & cm0 & 347.74 \\ \hline
5 & 6.06 & 4 & 0.25 & cm0, 5, 7, 15 & 0, 267.65, 0, 0.01 \\ \hline
6 & 6.05 & 5 & 0.25 & cm0, 7, 8, 10, 11 & 269.86, 0, 0, 0, 0 \\ \hline
7 & 6.03 & 4 & 0.24 & cm0, 5, 12, 18 & 278.62, 0, 0, 0 \\ \hline
8 & 4.85 & 5 & 0.21 & cm5, 6, 7, 11, 12 & 257.88, 0, 0, 0, 0 \\ \hline
9 & 4.19 & 4 & 0.2 & cm2, 12, 13, 18 & 230.49, 0.48, 0, 0 \\ \hline
10 & 4.11 & 1 & 0.18 & cm5 & 250.77 \\ \hline
11 & 4.07 & 3 & 0.24 & cm0, 6, 8 & 187.74, 0, 0 \\ \hline
12 & 3.94 & 4 & 0.22 & cm3, 5, 6, 13 & 0, 0, 197.59, 0 \\ \hline
13 & 3.93 & 2 & 0.2 & cm2, 8 & 214.58, 0 \\ \hline
14 & 3.82 & 1 & 0.21 & cm0 & 197.11 \\ \hline
15 & 3.7 & 1 & 0.21 & cm0 & 191.08 \\ \hline
16 & 3.66 & 8 & 0.32 & cm0, 2, 5, 10, 11, 12, 14, 16 & 127.07, 0, 0, 0, 0, 0, 0, 0 \\ \hline
17 & 3.65 & 4 & 0.23 & cm0, 9, 10, 18 & 176.98, 0, 0, 0 \\ \hline
18 & 3.5 & 2 & 0.21 & cm0, 16 & 179.86, 0 \\ \hline
19 & 3.48 & 2 & 0.23 & cm0, 7 & 170.21, 0 \\ \hline
20 & 3.48 & 3 & 0.23 & cm0, 6, 15 & 166.23, 0, 0 \\ \hline
\end{tabular}
\end{adjustbox}
\end{center}
From CM1 viewpoint (i.e. with CM1 in the role of the reference clearing member), 15 scenarios entail significant losses over the collateral posted by the defaulted CM0 (positive first entries in the last column of Table \ref{tab:RSTscDetailsexample}). CM0 bears a very large concentrated position compared to other members. Even if CM0 has more IM and DF requirements than others, this is still not enough: this example highlights that employed DF allocation rules in this example dilute the DF collateral requirements for concentrated positions. 
It also illustrates that scenarios with multiple defaults do not necessarily lead to extreme losses, due to the fact that members with medium or small positions have large default fund contributions stemming from others' concentrated positions.

In \hyperref[tab:STexo2CCPscase]{Table \ref{tab:STexo2CCPscase}}, we report, for the example illustrated by \hyperref[fig:Network2CCPsUC]{Figure \ref{fig:Network2CCPsUC}} with 2 CCPs, the trading loss levels (value-at-risks) at confidence levels $90\%$ and $99.9\%$, for the 24 common members on the two CCPs. The corresponding numbers in the case where the two CCPs would be considered separately is reported in the columns labeled ``stand-alone''.
\begin{center}
\begin{adjustbox}{
addcode={
\begin{minipage}{\width}}
{\caption{Quantile loss levels (confidence errors) for $90\%$ and $99.9\%$ confidence levels  across members for the example with 2 CCPs and 155 members including 24 common members. Legend for column headers: I. Member Id, II. DP (\%), III. Size on CCP1, IV. Volatility on CCP1, V. Size on CCP2, VI. Volatility on CCP2, VII. $90^{th}$ Perc. stand-alone, VIII. $90^{th}$ Perc., IX. $99.9^{th}$ Perc. stand-alone, $99.9^{th}$ Perc.}\label{tab:STexo2CCPscase}
\end{minipage}},
rotate=90,
width=0.9
\textwidth,
center=\textwidth,
float=table
}
\begin{tabular}{ |c|c|c|c|c|c|c|c|c|c| }
\hline
I & II & III & IV & V & VI & VII & VIII & IX & X \\ \hline
3 & 0.1 & 19.9 & 21 & -97.48 & 23 & 0.0227 (-3\%, 2.8\%) & 0.0442 (-1.3\%, 1.2\%) & 3.0505 (-2.2\%, 1.9\%) & 2.6768 (-2\%, 2.3\%) \\ \hline 
4 & 0.1 & 80.79 & 24 & -18.79 & 22 & 0.0215 (-2.7\%, 2.8\%) & 0.0428 (-1.1\%, 1.1\%) & 2.6739 (-2\%, 1.8\%) & 2.3058 (-2\%, 2.2\%) \\ \hline 
9 & 3.1 & -31.58 & 29 & 17.74 & 23 & 0.0195 (-1.9\%, 1.8\%) & 0.0399 (-0.9\%, 0.9\%) & 1.631 (-2.1\%, 2.3\%) & 1.2558 (-2.3\%, 1.9\%) \\ \hline 
12 & 0.1 & 17.97 & 21 & -16.75 & 24 & 0.0175 (-1.3\%, 1.2\%) & 0.0288 (-0.7\%, 0.7\%) & 0.8226 (-1.8\%, 2\%) & 0.5575 (-1.7\%, 2.1\%) \\ \hline 
13 & 0.1 & -14.9 & 22 & 15.81 & 25 & 0.0172 (-1.2\%, 1.2\%) & 0.0274 (-0.7\%, 0.7\%) & 0.7568 (-1.8\%, 2\%) & 0.5124 (-1.8\%, 1.8\%) \\ \hline 
14 & 0.2 & 12.34 & 23 & -14.93 & 26 & 0.0168 (-1.2\%, 1.2\%) & 0.0261 (-0.7\%, 0.7\%) & 0.7005 (-1.8\%, 2\%) & 0.4763 (-1.6\%, 1.6\%) \\ \hline 
15 & 0.1 & -10.23 & 24 & 14.09 & 27 & 0.0164 (-1.1\%, 1.1\%) & 0.0246 (-0.7\%, 0.7\%) & 0.6451 (-1.7\%, 2\%) & 0.4426 (-1.6\%, 1.5\%) \\ \hline 
17 & 0.3 & -7.03 & 26 & -13.3 & 28 & 0.016 (-1.1\%, 1.1\%) & 0.0226 (-0.7\%, 0.7\%) & 0.5674 (-1.7\%, 2\%) & 0.3998 (-1.6\%, 1.9\%) \\ \hline 
19 & 0.2 & -4.83 & 28 & 12.56 & 29 & 0.0155 (-1.1\%, 1\%) & 0.0205 (-0.7\%, 0.7\%) & 0.4998 (-1.6\%, 2\%) & 0.3673 (-1.6\%, 1.5\%) \\ \hline 
22 & 3.9 & 2.75 & 20 & -11.86 & 30 & 0.0173 (-1\%, 1\%) & 0.0197 (-0.8\%, 0.8\%) & 0.4577 (-1.8\%, 1.7\%) & 0.3856 (-1.4\%, 2.1\%) \\ \hline 
26 & 0.1 & 1.3 & 24 & 11.2 & 20 & 0.0094 (-0.9\%, 0.9\%) & 0.0104 (-0.8\%, 0.8\%) & 0.2448 (-1.6\%, 2\%) & 0.2099 (-1.6\%, 1.8\%) \\ \hline 
27 & 0.1 & 1.07 & 25 & -10.57 & 21 & 0.0093 (-0.9\%, 0.9\%) & 0.0101 (-0.7\%, 0.8\%) & 0.2377 (-1.6\%, 2\%) & 0.2074 (-1.5\%, 2\%) \\ \hline 
28 & 1.5 & 0.89 & 26 & 9.98 & 22 & 0.0097 (-0.9\%, 0.9\%) & 0.0104 (-0.8\%, 0.8\%) & 0.2433 (-2.1\%, 1.6\%) & 0.215 (-1.8\%, 1.8\%) \\ \hline 
31 & 0.1 & -0.51 & 29 & -9.42 & 23 & 0.009 (-0.9\%, 0.9\%) & 0.0094 (-0.8\%, 0.8\%) & 0.2191 (-1.7\%, 1.8\%) & 0.2021 (-1.5\%, 1.7\%) \\ \hline 
34 & 0.1 & 0.29 & 21 & 8.89 & 24 & 0.0089 (-0.9\%, 0.8\%) & 0.009 (-0.8\%, 0.9\%) & 0.2051 (-1.5\%, 2\%) & 0.1979 (-1.4\%, 2\%) \\ \hline 
35 & 0.1 & -0.24 & 22 & -8.4 & 25 & 0.0087 (-0.9\%, 0.9\%) & 0.0088 (-0.9\%, 0.8\%) & 0.201 (-1.5\%, 2\%) & 0.1948 (-1.4\%, 2\%) \\ \hline 
36 & 0.1 & 0.2 & 23 & 7.93 & 26 & 0.0086 (-0.9\%, 0.8\%) & 0.0086 (-0.8\%, 0.9\%) & 0.197 (-1.6\%, 1.9\%) & 0.1916 (-1.5\%, 1.8\%) \\ \hline 
39 & 0.1 & -0.11 & 26 & -7.48 & 27 & 0.0084 (-0.9\%, 0.8\%) & 0.0084 (-0.8\%, 0.9\%) & 0.1907 (-1.5\%, 2\%) & 0.1873 (-1.4\%, 2\%) \\ \hline 
40 & 0.5 & 0.09 & 27 & 7.07 & 28 & 0.0084 (-0.8\%, 0.9\%) & 0.0084 (-0.8\%, 0.9\%) & 0.1898 (-1.5\%, 2\%) & 0.1869 (-1.4\%, 2\%) \\ \hline 
44 & 0.1 & 0.04 & 20 & -6.67 & 29 & 0.008 (-0.8\%, 0.8\%) & 0.008 (-0.8\%, 0.8\%) & 0.1806 (-1.5\%, 1.9\%) & 0.1796 (-1.5\%, 1.9\%) \\ \hline 
49 & 0.1 & -0.02 & 25 & 6.3 & 30 & 0.0078 (-0.9\%, 0.8\%) & 0.0079 (-0.8\%, 0.8\%) & 0.1758 (-1.4\%, 2\%) & 0.1752 (-1.4\%, 2\%) \\ \hline 
50 & 0.1 & 0.01 & 26 & -5.95 & 20 & 0.0049 (-0.9\%, 0.8\%) & 0.0049 (-0.8\%, 0.8\%) & 0.1109 (-1.6\%, 1.9\%) & 0.1106 (-1.6\%, 1.9\%) \\ \hline 
51 & 0.1 & -0.01 & 27 & 5.61 & 21 & 0.0049 (-0.8\%, 0.8\%) & 0.0049 (-0.8\%, 0.8\%) & 0.1095 (-1.5\%, 2\%) & 0.1092 (-1.5\%, 2\%) \\ \hline 
55 & 0.1 & -0.01 & 20 & -5.3 & 22 & 0.0048 (-0.9\%, 0.8\%) & 0.0048 (-0.8\%, 0.8\%) & 0.1084 (-1.5\%, 1.9\%) & 0.1082 (-1.5\%, 1.9\%) \\ \hline 
\end{tabular}
\end{adjustbox}
\end{center}
For quantiles at $90\%$ confidence levels, the loss levels are significantly higher when the common membership are considered compared to the stand-alone quantile loss calculation conducted on each CCP and summed, especially for the first ten members.

For members with very low size on one of the two CCPs compared to the other, considering the common memberships or not does not affect the loss estimates, as expected\footnote{as the CCP with the very low size compared to the other should have marginal impact.}. This outlines the importance of taking into account such commonality feature for sizeable members on the CCPs. On the contrary, with quantile loss levels at confidence level $99.9\%$, the sum of stand-alone loss estimations are well above the loss estimate when common memberships are taken into consideration. For members facing the two CCPs, this leads in particular to over-conservative KVA estimates. This in turn is detrimental for client end-users that support unnecessary additional capital costs.

\section{Optimizing the \Novation of Defaulted Client Portfolios}\label{sec:OptNovDefMbb}
 
In case a clearing member defaults, the 
CCP tentatively novates part of the CCP portfolio of the defaulted member through auctions among the surviving clearing members \citep{Hoogland2000,BIS-auctions-19}, and it liquidates the residual on the market. A natural baseline is that the CCP novates (auctions among surviving members) client trades and their mirroring client account positions, collectively dubbed client positions for brevity hereafter, whereas house account positions are liquidated. 

The liquidation side of the procedure cannot be handled in our modeling setup,
which does not embed the fundamentals of price formation (our MtM processes are assumed to be exogenously given). 
On the other hand, an XVA-based procedure can be used for rendering what would be the output of an idealized, efficient auction, assuming a large number of clearing members \citep[Section 3.3]{oleschak2019central}.
Namely, supposing that the reference clearing member, labeled by 0 in Sections \ref{s:setup}-\ref{s:xvaa}, defaults at time 0, i.e. just after that all portfolios have been settled, for each surviving member $\CM *$ successively envisioned as a potential taker of the defaulted (client) positions of CM0, one computes the incremental ($\Delta$) XVAs of porting 
the defaulted positions to $\CM *$, for each surviving member ($\CM *$ included\footnote{note that all members are impacted by additional margin to fund due to the re-calibration of their DF by the CCP, whereas only the member taker of the portfolio sees in addition its IM adjusted.}). The corresponding incremental XVA numbers are then summed over metrics and survivors, resulting in 
the funds transfer price ($\FTP *$) of porting defaulted client positions to $\CM *$. The effective taker is then the surviving member for which the ensuing $\FTP *$ is the smallest\footnote{or, indifferently in case of multiple minima, one of the minimizing $\FTP *$ members.}. See   \citeN[Section 5.2]{AlbaneseChataignerCrepey20} for more details on such ``XVA Pareto optimally" driven novation procedures.

In what follows, based on the example of \hyperref[tab:MbNework1CCP20MbSetup]{Table \ref{tab:MbNework1CCP20MbSetup}} (which only involves client positions), we analyze from this perspective a first scenario of a single default on the CCP and a second scenario with two defaults.

\subsection{Single Default Resolution Example}

Taking the first case with a single default, we first assume the scenario whereby CM0 defaults at time $0$.  
\hyperref[tab:FTPXVAsCosts]{Table \ref{tab:FTPXVAsCosts}} summarizes the $\Delta\XVA*$, across members $*$ from $1$ to $19$, in increasing order of the total $\FTP*$ indicated in the last column.  
Based on the results of \hyperref[tab:FTPXVAsCosts]{Table \ref{tab:FTPXVAsCosts}}, CM1 appears to be the potential
taker leading to the least overall FTP costs across all surviving members. This is understandable as this member's portfolio size (184 in Table  \ref{tab:MbNework1CCP20MbSetup}) nets the most the defaulted member's portfolio size (-242), with volatility and credit default probability similar to\footnote{in particular, not significantly higher than.}  
the ones of the defaulted member.
\begin{center}
\begin{adjustbox}{width=\textwidth,center=\textwidth,captionbelow={$\Delta\XVA*$ corresponding to the different surviving $\CM*$, i.e. for $*$ other than 0, assuming an instant default of CM0 at time 0. In parenthesis, the contributions to $\Delta\XVA*$ of $\CM*$ itself.},label={tab:FTPXVAsCosts},float=table}
\begin{tabular}{ |c|c|c|c|c| }
\hline
Surv. member $*$ $-$ Costs & Total $\Delta\CMVA*$ & Total $\Delta\CCVA* $ & Total $\Delta\KVA*$ & Total $\FTP*$\\ \hline
1 & 0.0768 (0.0295) & -0.0025 (0.0018) & 0.0743 (-0.0006) & 0.1486 (0.0307)\\ \hline
2 & 0.0921 (0.0428) & 0.0007 (0.0047) & 0.1418 (0.0387) & 0.2346 (0.0862)\\ \hline
19 & 0.1298 (0.0818) & 0.0129 (0.0223) & 0.1725 (0.2313) & 0.3151 (0.3354)\\ \hline
3 & 0.1054 (0.0576) & 0.007 (0.0075) & 0.2333 (0.0735) & 0.3457 (0.1387)\\ \hline
18 & 0.1417 (0.0939) & 0.0186 (0.0219) & 0.2757 (0.2275) & 0.4359 (0.3432)\\ \hline
17 & 0.1549 (0.107) & 0.0228 (0.0218) & 0.3215 (0.229) & 0.4992 (0.3578)\\ \hline
16 & 0.1688 (0.1208) & 0.0271 (0.0217) & 0.4075 (0.2303) & 0.6034 (0.3728)\\ \hline
4 & 0.1525 (0.1022) & 0.0116 (0.0122) & 0.445 (0.1288) & 0.6091 (0.2431)\\ \hline
15 & 0.1814 (0.1334) & 0.0305 (0.0214) & 0.4656 (0.2294) & 0.6775 (0.3842)\\ \hline
14 & 0.1903 (0.1426) & 0.035 (0.0209) & 0.5079 (0.2237) & 0.7332 (0.3872)\\ \hline
13 & 0.2061 (0.1582) & 0.0384 (0.0208) & 0.5903 (0.2272) & 0.8349 (0.4063)\\ \hline
12 & 0.2171 (0.1692) & 0.0406 (0.0202) & 0.6277 (0.2252) & 0.8854 (0.4147)\\ \hline
11 & 0.2285 (0.1807) & 0.0439 (0.0198) & 0.6811 (0.2223) & 0.9536 (0.4228)\\ \hline
10 & 0.2385 (0.1908) & 0.0469 (0.019) & 0.757 (0.2181) & 1.0424 (0.428)\\ \hline
8 & 0.234 (0.1881) & 0.0512 (0.0164) & 0.7812 (0.1858) & 1.0663 (0.3903)\\ \hline
7 & 0.2327 (0.1876) & 0.0519 (0.0149) & 0.809 (0.1696) & 1.0936 (0.3721)\\ \hline
9 & 0.2478 (0.2003) & 0.0483 (0.0181) & 0.7994 (0.2117) & 1.0955 (0.4301)\\ \hline
6 & 0.2687 (0.2225) & 0.0506 (0.0135) & 0.9811 (0.1689) & 1.3004 (0.405)\\ \hline
5 & 0.2728 (0.2274) & 0.0486 (0.0113) & 1.0242 (0.1414) & 1.3456 (0.3801)\\ \hline
\end{tabular} 
\end{adjustbox}
\end{center}

As CM1 concentrates more risks due in particular to non-perfect offset\footnote{By offset we refer to risk reduction when taking over some additional position. The effect of correlation is such that an opposite sign in portfolio size does not imply an equal offset of the risk of the aggregated positions. For instance, even with opposite sizes and same volatilities but for $\rho^{mkt}\in(0,1/2)$, the member ends up with more risk.}
 between its prior positions and the defaulting one, there is an increase of its IM reflected through an increase of CMVA. But the new risk of CM1 is less than the sum of the former risks of CM0 and CM1, hence the CCVA aggregated across surviving members is reduced. This only happens when CM1 takes over the defaulting portfolio, other potential takers leading to an overall increase of the CCVA. As for the KVA, there is a reduction effect for CM1 when CM1 is the taker (see the term in parentheses in Table \ref{tab:FTPXVAsCosts}), but an overall increase in the total KVA (aggregated over all surviving members), whichever the member taker is. Having CM1 as the taker allows to minimize $\Delta\KVA$.\smallskip

As expected, among the three XVA components, KVA is the main determinant of the optimal taker: see  \hyperref[tab:AvAbsDist1CCPcase]{Table \ref{tab:AvAbsDist1CCPcase}}.
\begin{center}
\begin{adjustbox}{width=0.35\textwidth,center=\textwidth,captionbelow={
Standard deviation across surviving members $*$ of the $\Delta\XVA*$'s
for the example with 1 CCP and 20 members, assuming an instant default of CM0 at time 0.},label={tab:AvAbsDist1CCPcase},float=table}
\begin{tabular}{ ccc }
\hline
$\Delta \CMVA$ & $\Delta\CCVA $ & $\Delta \KVA$ \\ \hline\hline
0.0593 & 0.0182 & 0.2846\\ \hline
\end{tabular}
\end{adjustbox}
\end{center}

\begin{figure}
\begin{center}
\begin{tikzpicture}[
roundnodeC/.style={circle, draw=green!60, fill=green!5, thick, minimum size=5mm},
roundnodeB/.style={circle, draw=blue!60, fill=blue!5, very thick, minimum size=7mm},
roundnodeCH/.style={circle, draw=red!60, fill=red!5, very thick, minimum size=15mm},
scale=0.94, transform shape
]

%Nodes
\node at (0,0) [roundnodeCH]  (theCCP)    {CCP};
\node at (-3,0) [roundnodeB, dashed, opacity=.4]   (bank1)     {\scriptsize B0};
\node at (-2.8,0.77) [roundnodeB]   (bank2)     {\scriptsize B1};
\node at (-2.3,1.4) [roundnodeB]   (bank3)     {\scriptsize B2};
\node at (-1.57,1.75) [roundnodeB]   (bank4)     {\scriptsize B3};
\node at (-0.8,1.95) [roundnodeB]   (bank5)     {\scriptsize B4};
\node at (0,2) [roundnodeB]   (bank6)     {\scriptsize B5};
\node at (0.8,1.95) [roundnodeB]   (bank7)     {\scriptsize B6};
\node at (1.57,1.75) [roundnodeB]   (bank8)     {\scriptsize B7};
\node at (2.3,1.4) [roundnodeB]   (bank9)     {\scriptsize B8};
\node at (2.8,0.77) [roundnodeB]   (bank10)     {\tiny B9};
\node at (3,0) [roundnodeB]   (bank11)     {\tiny B10};
\node at (2.8,-0.77) [roundnodeB]   (bank12)     {\tiny B11};
\node at (2.3,-1.4) [roundnodeB]   (bank13)     {\tiny B12};
\node at (1.57,-1.75) [roundnodeB]   (bank14)     {\tiny B13};
\node at (0.8,-1.95) [roundnodeB]   (bank15)     {\tiny B14};
\node at (0,-2) [roundnodeB]   (bank16)     {\tiny B15};
\node at (-0.8,-1.95) [roundnodeB]   (bank17)     {\tiny B16};
\node at (-1.57,-1.75) [roundnodeB]   (bank18)     {\tiny B17};
\node at (-2.3,-1.4) [roundnodeB]   (bank19)     {\tiny B18};
\node at (-2.8,-0.77) [roundnodeB]   (bank20)     {\tiny B19};

\node at (0,-3.4) [roundnodeC]  (client16)    {\scriptsize C15};

\node at (0,3.4) [roundnodeC]  (client6)    {\scriptsize C5};
\node at (-1.1,3.3) [roundnodeC]  (client5)    {\scriptsize C4};
\node at (-2.2,3) [roundnodeC]  (client4)    {\scriptsize C3};
\node at (-3.4,2.5) [roundnodeC]  (client3)    {\tiny C2};
\node at (-4.4,1.7) [roundnodeC]  (client2)    {\tiny C1};
\node at (-4.8,0.5) [roundnodeC]  (client1)    {\tiny C0};
\node at (1.1,3.3) [roundnodeC]  (client7)    {\tiny C6};
\node at (2.2,3) [roundnodeC]  (client8)    {\tiny C7};
\node at (3.4,2.5) [roundnodeC]  (client9)    {\tiny C8};
\node at (4.4,1.7) [roundnodeC]  (client10)    {\tiny C9};
\node at (5,0) [roundnodeC]  (client11)    {\tiny C10};

\node at (1.1,-3.3) [roundnodeC]  (client15)    {\tiny C14};
\node at (2.2,-3) [roundnodeC]  (client14)    {\tiny C13};
\node at (3.4,-2.5) [roundnodeC]  (client13)    {\tiny C12};
\node at (4.4,-1.7) [roundnodeC]  (client12)    {\tiny C11};

\node at (-1.1,-3.3) [roundnodeC]  (client17)    {\tiny C16};
\node at (-2.2,-3) [roundnodeC]  (client18)    {\tiny C17};
\node at (-3.4,-2.5) [roundnodeC]  (client19)    {\tiny C18};
\node at (-4.4,-1.7) [roundnodeC]  (client20)    {\tiny C19};

%Lines
\draw[-,opacity=.4,dashed] (bank1.east) -- (theCCP.west);
\draw[-] (bank2.-10) -- (theCCP.165);
\draw[-,line width=.5mm] (bank2.-25) -- (theCCP.173);
\draw[-] (bank3.-45) -- (theCCP.150);
\draw[-] (bank4.-45) -- (theCCP.130);
\draw[-] (bank5.-67) -- (theCCP.110);
\draw[-] (bank6.south) -- (theCCP.north);
\draw[-] (bank7.-113) -- (theCCP.70);
\draw[-] (bank8.-135) -- (theCCP.50);
\draw[-] (bank9.-135) -- (theCCP.30);
\draw[-] (bank10.-158) -- (theCCP.10);
\draw[-] (bank11.west) -- (theCCP.east);
\draw[-] (bank12.158) -- (theCCP.-10);
\draw[-] (bank13.135) -- (theCCP.-30);
\draw[-] (bank14.135) -- (theCCP.-50);
\draw[-] (bank15.113) -- (theCCP.-70);
\draw[-] (bank16.north) -- (theCCP.south);
\draw[-] (bank17.67) -- (theCCP.-110);
\draw[-] (bank18.45) -- (theCCP.-130);
\draw[-] (bank19.45) -- (theCCP.-150);
\draw[-] (bank20.22) -- (theCCP.-170);

\draw[-,line width=0.5mm] (bank2.west) -- (client1.10);
\draw[-, dashed, opacity=.4] (bank1.170) -- (client1.-10);
\draw[-] (bank2.158) -- (client2.-45);
\draw[-] (bank3.135) -- (client3.-45);
\draw[-] (bank4.113) -- (client4.-67);
\draw[-] (bank5.113) -- (client5.south);
\draw[-] (bank6.north) -- (client6.south);
\draw[-] (bank10.22) -- (client10.-135);
\draw[-] (bank9.45) -- (client9.-135);
\draw[-] (bank8.67) -- (client8.-113);
\draw[-] (bank7.67) -- (client7.south);
\draw[-] (bank11.east) -- (client11.west);
\draw[-] (bank20.-158) -- (client20.45);
\draw[-] (bank19.-135) -- (client19.45);
\draw[-] (bank18.-113) -- (client18.67);
\draw[-] (bank17.-113) -- (client17.north);
\draw[-] (bank16.south) -- (client16.north);
\draw[-] (bank15.-67) -- (client15.north);
\draw[-] (bank14.-67) -- (client14.113);
\draw[-] (bank13.-45) -- (client13.135);
\draw[-] (bank12.-22) -- (client12.135);
\end{tikzpicture}
\end{center}
\caption{The 1-CCP, former 20-member financial network with 19 members post CM0 default. Defaulted CM0, labeled ``B0'' in the presented network, is represented as pale dashed node with pale dashed links to reflect former exposures to its client and toward the CCP. The optimal \novation  of CM0 portfolio with CM1, labeled ``B1'', is outlined with bold links to reflect the new exposures for CM1.}
\label{fig:Network1CCP19MbsNov}
\end{figure}
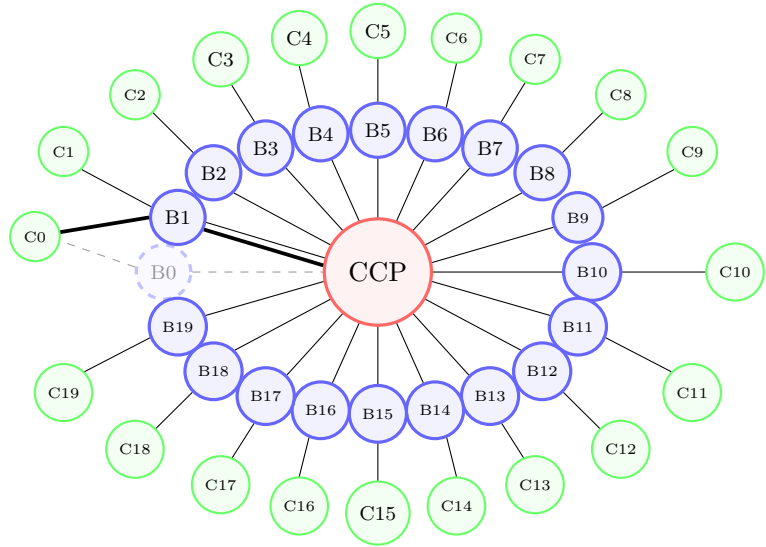

Once the CCP has re-allocated all defaulted client positions, the resulting financial network formerly depicted in Figure \ref{fig:Network1CCP20Mbs} becomes the network with 19 members shown in Figure \ref{fig:Network1CCP19MbsNov}. The thick lines represent the new portfolio exposures for CM1 and the pale dashed lines show the defaulted CM0 positions.

\subsection{Joint Default Resolution Example}

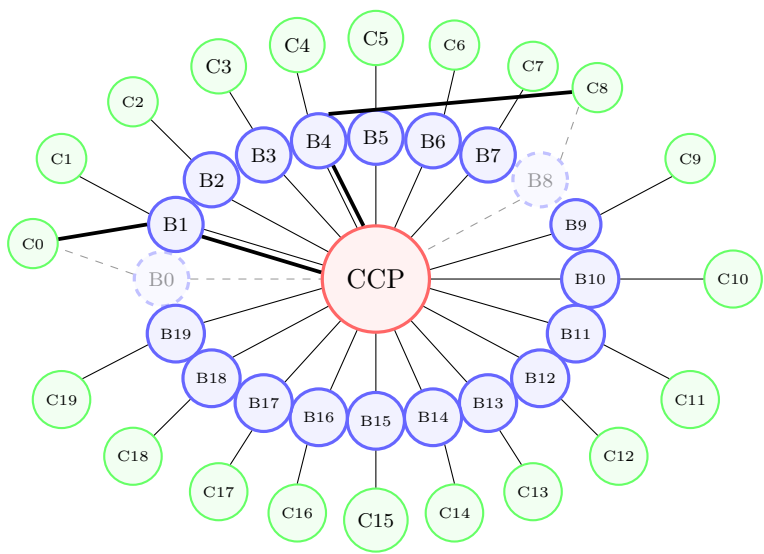
\begin{figure}[ht]
\begin{center}
\begin{tikzpicture}[
roundnodeC/.style={circle, draw=green!60, fill=green!5, thick, minimum size=5mm},
roundnodeB/.style={circle, draw=blue!60, fill=blue!5, very thick, minimum size=7mm},
roundnodeCH/.style={circle, draw=red!60, fill=red!5, very thick, minimum size=15mm},
scale=0.94, transform shape
]

%Nodes
\node at (0,0) [roundnodeCH]  (theCCP)    {CCP};
\node at (-3,0) [roundnodeB, dashed, opacity=.4]   (bank1)     {\scriptsize B0};
\node at (-2.8,0.77) [roundnodeB]   (bank2)     {\scriptsize B1};
\node at (-2.3,1.4) [roundnodeB]   (bank3)     {\scriptsize B2};
\node at (-1.57,1.75) [roundnodeB]   (bank4)     {\scriptsize B3};
\node at (-0.8,1.95) [roundnodeB]   (bank5)     {\scriptsize B4};
\node at (0,2) [roundnodeB]   (bank6)     {\scriptsize B5};
\node at (0.8,1.95) [roundnodeB]   (bank7)     {\scriptsize B6};
\node at (1.57,1.75) [roundnodeB]   (bank8)     {\scriptsize B7};
\node at (2.3,1.4) [roundnodeB, dashed, opacity=.4]   (bank9) {\scriptsize B8};
\node at (2.8,0.77) [roundnodeB]   (bank10)     {\tiny B9};
\node at (3,0) [roundnodeB]   (bank11)     {\tiny B10};
\node at (2.8,-0.77) [roundnodeB]   (bank12)     {\tiny B11};
\node at (2.3,-1.4) [roundnodeB]   (bank13)     {\tiny B12};
\node at (1.57,-1.75) [roundnodeB]   (bank14)     {\tiny B13};
\node at (0.8,-1.95) [roundnodeB]   (bank15)     {\tiny B14};
\node at (0,-2) [roundnodeB]   (bank16)     {\tiny B15};
\node at (-0.8,-1.95) [roundnodeB]   (bank17)     {\tiny B16};
\node at (-1.57,-1.75) [roundnodeB]   (bank18)     {\tiny B17};
\node at (-2.3,-1.4) [roundnodeB]   (bank19)     {\tiny B18};
\node at (-2.8,-0.77) [roundnodeB]   (bank20)     {\tiny B19};

\node at (0,-3.4) [roundnodeC]  (client16)    {\scriptsize C15};

\node at (0,3.4) [roundnodeC]  (client6)    {\scriptsize C5};
\node at (-1.1,3.3) [roundnodeC]  (client5)    {\scriptsize C4};
\node at (-2.2,3) [roundnodeC]  (client4)    {\scriptsize C3};
\node at (-3.4,2.5) [roundnodeC]  (client3)    {\tiny C2};
\node at (-4.4,1.7) [roundnodeC]  (client2)    {\tiny C1};
\node at (-4.8,0.5) [roundnodeC]  (client1)    {\tiny C0};
%\node at (-5,0.5) [roundnodeC]  (client1)    {\tiny C1};
%\node at (-5,-0.5) [roundnodeC]  (client0)    {\tiny C0};
\node at (1.1,3.3) [roundnodeC]  (client7)    {\tiny C6};
\node at (2.2,3) [roundnodeC]  (client8)    {\tiny C7};
\node at (3.1,2.7) [roundnodeC]  (client9)    {\tiny C8};
\node at (4.4,1.7) [roundnodeC]  (client10)    {\tiny C9};
\node at (5,0) [roundnodeC]  (client11)    {\tiny C10};

\node at (1.1,-3.3) [roundnodeC]  (client15)    {\tiny C14};
\node at (2.2,-3) [roundnodeC]  (client14)    {\tiny C13};
\node at (3.4,-2.5) [roundnodeC]  (client13)    {\tiny C12};
\node at (4.4,-1.7) [roundnodeC]  (client12)    {\tiny C11};

\node at (-1.1,-3.3) [roundnodeC]  (client17)    {\tiny C16};
\node at (-2.2,-3) [roundnodeC]  (client18)    {\tiny C17};
\node at (-3.4,-2.5) [roundnodeC]  (client19)    {\tiny C18};
\node at (-4.4,-1.7) [roundnodeC]  (client20)    {\tiny C19};

%Lines
\draw[-,opacity=.4,dashed] (bank1.east) -- (theCCP.west);
\draw[-] (bank2.-10) -- (theCCP.165);
\draw[-,line width=.5mm] (bank2.-25) -- (theCCP.173);
\draw[-] (bank3.-45) -- (theCCP.150);
\draw[-] (bank4.-45) -- (theCCP.130);
\draw[-] (bank5.-72) -- (theCCP.110);
\draw[-,line width=.5mm] (bank5.-60) -- (theCCP.103);
\draw[-] (bank6.south) -- (theCCP.north);
\draw[-] (bank7.-113) -- (theCCP.70);
\draw[-] (bank8.-135) -- (theCCP.50);
\draw[-,opacity=.4,dashed] (bank9.-135) -- (theCCP.30);
\draw[-] (bank10.-158) -- (theCCP.10);
\draw[-] (bank11.west) -- (theCCP.east);
\draw[-] (bank12.158) -- (theCCP.-10);
\draw[-] (bank13.135) -- (theCCP.-30);
\draw[-] (bank14.135) -- (theCCP.-50);
\draw[-] (bank15.113) -- (theCCP.-70);
\draw[-] (bank16.north) -- (theCCP.south);
\draw[-] (bank17.67) -- (theCCP.-110);
\draw[-] (bank18.45) -- (theCCP.-130);
\draw[-] (bank19.45) -- (theCCP.-150);
\draw[-] (bank20.22) -- (theCCP.-170);

\draw[-,line width=0.5mm] (bank2.west) -- (client1.10);
\draw[-, dashed, opacity=.4] (bank1.170) -- (client1.-10);
\draw[-] (bank2.158) -- (client2.-45);
\draw[-] (bank3.135) -- (client3.-45);
\draw[-] (bank4.113) -- (client4.-67);
\draw[-,line width=0.5mm] (bank5.70) -- (client9.-170);
\draw[-] (bank5.113) -- (client5.south);
\draw[-] (bank6.north) -- (client6.south);
\draw[-] (bank10.22) -- (client10.-135);
\draw[-, dashed, opacity=.4] (bank9.45) -- (client9.-135);
\draw[-] (bank8.67) -- (client8.-113);
\draw[-] (bank7.67) -- (client7.south);
\draw[-] (bank11.east) -- (client11.west);
\draw[-] (bank20.-158) -- (client20.45);
\draw[-] (bank19.-135) -- (client19.45);
\draw[-] (bank18.-113) -- (client18.67);
\draw[-] (bank17.-113) -- (client17.north);
\draw[-] (bank16.south) -- (client16.north);
\draw[-] (bank15.-67) -- (client15.north);
\draw[-] (bank14.-67) -- (client14.113);
\draw[-] (bank13.-45) -- (client13.135);
\draw[-] (bank12.-22) -- (client12.135);
\end{tikzpicture}
\end{center}
\caption{The 1-CCP, former 20 member financial network, with 18 members post CM0 and CM8 defaults. Defaulted CM0, labeled ``B0'' and CM8, labeled ``B8'', are represented as pale dashed node with pale dashed links to reflect former exposures to their clients and toward the CCP. The optimal \novations of CM0 and CM8 portfolios are outlined with bold links to reflect the new exposures for both CM1 and CM4.}
\label{fig:Network1CCP18MbsNov}
\end{figure}

In case two members default instantly at time 0,
it is likewise possible to resolve numerically the re-allocation of their client portfolios. The number of possible combinations of takers in that case is $18^2=324$ (assuming each of the two portfolios taken over by one survivor). By putting into default CM0, the largest member with portfolio size $-242$, as well as CM8, a middle-sized member with portfolio size $26$, we get that CM1 and CM4 taking over the respective portfolios of the defaulted CM0 and CM8 leads to the least FTP (additional $\Delta\XVA$s aggregated across the remaining 18 members of the CCP). The resulting network post defaults of members 0 and 8 is shown in Figure \ref{fig:Network1CCP18MbsNov}.
As depicted by \hyperref[tab:AvAbsDist1CCP2Defts]{Table \ref{tab:AvAbsDist1CCP2Defts}},
the KVA  again plays the major role in determining the optimal takers. 
 \begin{center}
\begin{adjustbox}{width=0.35\textwidth,center=\textwidth,captionbelow={Standard deviation of incremental XVAs across members for the example with 1 CCP and 20 members with CM0 and CM8 considered in default state},label={tab:AvAbsDist1CCP2Defts},float=table}
\begin{tabular}{ ccc }
\hline
$\Delta \CMVA$ & $\Delta\CCVA $ & $\Delta \KVA$ \\ \hline\hline
0.0586 & 0.0178 & 0.2875\\ \hline
\end{tabular}
\end{adjustbox}
\end{center}
When looking at the signed portfolio sizes and one-year default probability of the two takers, it is aligned with the intuition that the second largest member, which is CM1 with portfolio size $184$, should take over the defaulted portfolio of CM0  with size $-242$, as it has an opposite portfolio direction, resulting into a strong netting benefit. Moreover that its default probability is similar to that of CM1. At first sight, CM4 with prior default portfolio size $-80$ taking over defaulting portfolio of CM8 with size $26$ seems surprising. Other potential takers with closest opposite portfolios sizes are CM6 (with size $-46$) and CM9 (with size $-20$). But their default probability is roughly twice the one of CM4. As a result, CM4 taking over CM8 defaulted portfolio yields a significant reduction in terms of KVA compared to the situation where CM6 or CM9 would take over CM8's portfolio, as depicted in \hyperref[tab:DiffScFTPs2Deflts]{Table \ref{tab:DiffScFTPs2Deflts}}.
\begin{center}
\begin{adjustbox}{width=\textwidth,center=\textwidth,captionbelow={FTPs for different pairs of member takers on defaulted portfolio 0 and 8 for the case of 1 CCP and 20 members where information in \{ , \} relates to the first taking member on defaulted portfolio 0 and to the second taking member on defaulted portfolio 8.},label={tab:DiffScFTPs2Deflts},float=table}
\begin{tabular}{ cccccccc }
\hline
\{taker of 0, taker of 8\} & Sizes & Vol ( \% ) & DP ( \% ) & FTP MVA & FTP CCVA & FTP KVA & FTP\\ \hline\hline
\{ 1, 4 \} & \{ 184, -80 \} & \{ 21, 24 \} & \{ 0.6, 0.9 \} & 0.07358 & 0.00007 & 0.17576 & 0.24941\\ \hline
\{ 1, 6 \} & \{ 184, -46 \} & \{ 21, 26 \} & \{ 0.6, 1.9 \} & 0.0756 & 0.00255 & 0.18252 & 0.26067\\ \hline
\{ 1, 9 \} & \{ 184, -20 \} & \{ 21, 29 \} & \{ 0.6, 1.6 \} & 0.08394 & 0.00405 & 0.18827 & 0.27626\\ \hline
\end{tabular}
\end{adjustbox}
\end{center}

\begin{figure}[ht]
\begin{center}
\includegraphics[width=1\textwidth]{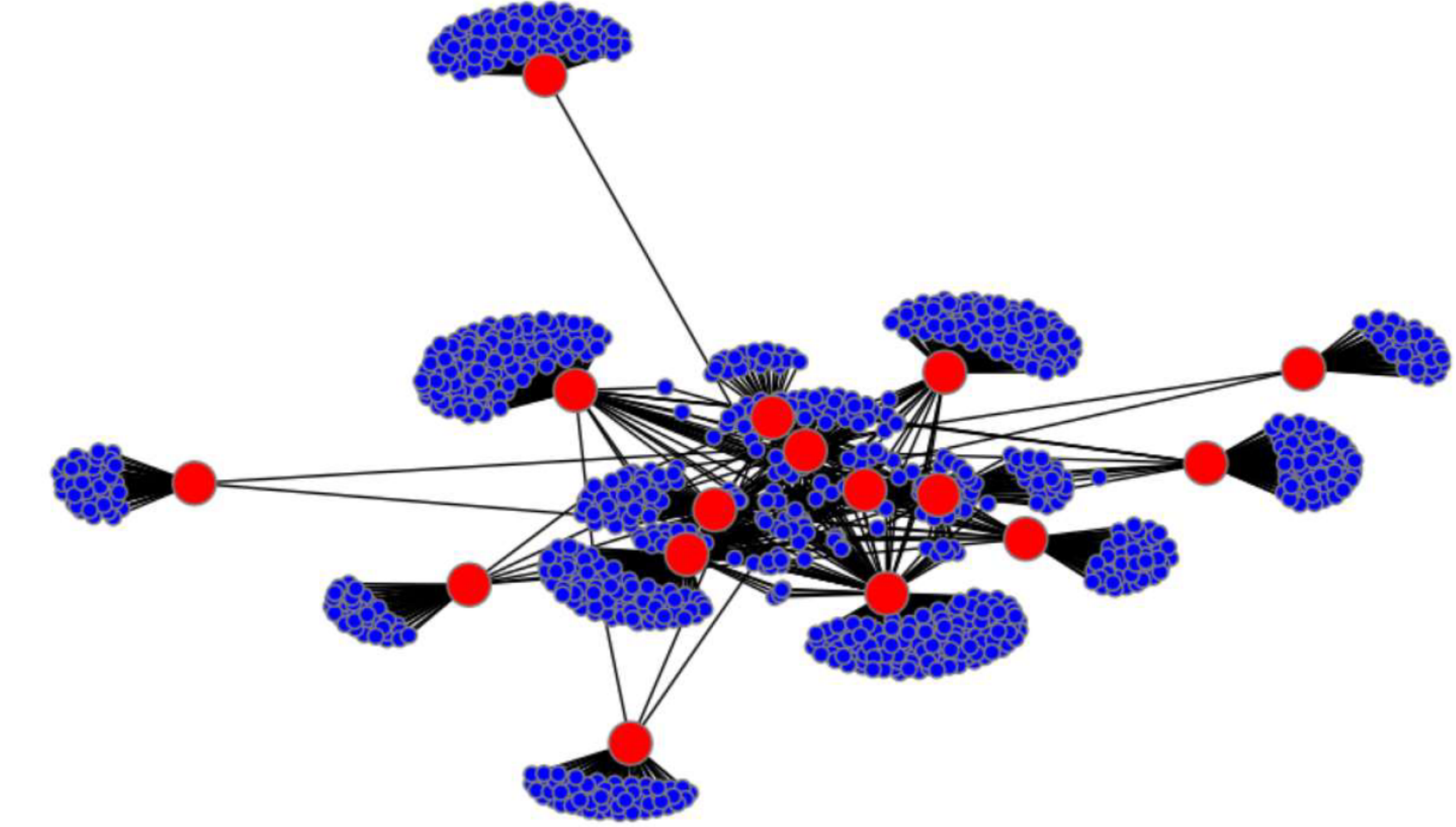}
\end{center}
\caption{Large clearing network example with 16 CCPs in red and their members in blue with many having common memberships concentrated in the center of the network.} 
\label{fig:LargeClearingNetworkExample}
\end{figure}

In practice, much larger financial networks are involved. An illustration of such network \b(restricted to Eurozone) is given by \hyperref[fig:LargeClearingNetworkExample]{Figure \ref{fig:LargeClearingNetworkExample}}, omitting all client trades for ease of readability. The center of the network indicates the various members having common memberships towards several CCPs.
Combinatorial novation optimization, as also stress test analysis\footnote{see Section \ref{ss:numres}}, over such complex networks, is of course order(s) of magnitude heavier than what we presented in the above and would require specialized numerical techniques.

\section{Conclusion\label{sec:concl}}

We have proposed a fully integrated risk management framework that can serve stress test analysis, including reverse stress test in line with regulatory requirements, as well as
porting defaulted portfolios analysis, in a setup encompassing all the trades (bilateral as centrally cleared and their hedges) of a reference bank. The framework includes dependence features between financial participants portfolios, joint defaults, and a configurable wrong-way risk feature. This is done in a numerically  tractable static setup (although already quite demanding on large financial networks). A dynamic extension could be considered (but at an even much higher computational burden).
Another improvement would be to add regulatory constraints such as minimum regulatory capital requirements and liquidity leverage ratios. More fundamentally, in this paper, we tackle the derivatives risk problem from a pure counterparty credit risk viewpoint: thus note that, if members, clients and counterparties are all default free, then in view of Proposition \ref{p:ccprm} all considered XVAs are zero and our setup trivializes. In fact, another dimension to the problem is liquidity (see e.g. \citeN{AminiFilipovicMinca19,FHT18}). Depending on the considered applications\footnote{see e.g. the beginning of Section \ref{sec:OptNovDefMbb}.}, credit or liquidity is the main force at hand. A challenging research project would be to integrate both in a common setup.

\bibliographystyle{chicago}
\bibliography{CCP-static}
\end{document}